\documentclass[lettersize,journal]{IEEEtran}
\usepackage{amsmath,amsfonts}
\usepackage{algorithmic}
\usepackage{algorithm}
\usepackage{array}
\usepackage[caption=false,font=footnotesize,labelfont=sf,textfont=sf]{subfig}
\usepackage{textcomp}
\usepackage{stfloats}
\usepackage{url}
\usepackage{verbatim}
\usepackage{graphicx}
\usepackage{cite}
\usepackage{tabularray}
\usepackage{multirow}
\usepackage{hhline}
\usepackage{colortbl}
\usepackage{amsmath}
\usepackage{color,xcolor}
\usepackage{bm}
\usepackage{upgreek}
\usepackage{tikz}
\usepackage{makecell}
\usepackage{hyperref}
\usepackage{pifont}
\usepackage{circledsteps}
\usepackage{tikz}
\usepackage{booktabs}
\usepackage{tabularx}
\usepackage{ragged2e}
\usepackage{color}

\usepackage{cleveref}
\crefname{figure}{Fig.}{Figs.}
\Crefname{figure}{Fig.}{Figs.}

\usepackage{orcidlink}
\usepackage{wasysym}
\usepackage{pifont}
\usepackage{mathptmx}
\DeclareMathAlphabet{\mathcal}{OMS}{cmsy}{m}{n}

\usepackage{tabu}
\usepackage{lipsum}
\usepackage{mwe}
\usepackage{float}
\usepackage{enumitem}
\usepackage[table]{xcolor}
\definecolor{a}{HTML}{B22222}
\definecolor{amber}{HTML}{FFD044}
\definecolor{orange}{HTML}{F2A46F}
\definecolor{burntorange}{HTML}{E65B00}

\newcommand{\circled}[1]{\tikz[baseline=(char.base)]{
    \node[shape=circle,draw,inner sep=0.5pt] (char) {\small #1};}}

\usepackage{CJKutf8}

\begin{document}
\bstctlcite{IEEEexample:BSTcontrol}

\title{MRATTS: An MR-Based Acupoint Therapy Training System with Real-Time Acupoint Detection and Evaluation Standards} 

\author{Jiacheng~Liu\orcidlink{0009-0000-9043-7393},
        Bohan~Chen\orcidlink{0009-0005-4171-5779},
        Qian~Wang\orcidlink{0009-0003-1213-5367},
        Weichao~Song\orcidlink{0009-0001-7972-4647},
        Fangfei~Ye\orcidlink{0009-0007-2178-8190},
        Liang~Zhou$^{*}$\orcidlink{0000-0002-0462-4131},
        Haibin~Ling$^{*}$\orcidlink{0000-0003-4094-8413},~\IEEEmembership{Fellow,~IEEE},
        and~Bingyao~Huang$^{*}$\orcidlink{0000-0002-8647-5730}
\thanks{Jiacheng Liu and Liang Zhou are with the Institute of Medical Technology, Peking University Health
Science Center, and the National Institute of Health Data Science, Peking University (e-mail: liujiacheng@stu.pku.edu.cn; zhoulng@pku.edu.cn). This work was partly done while Jiacheng Liu was with the College of Computer and Information Science, Southwest University.}
\thanks{Bohan Chen, Qian Wang and Bingyao Huang are with the College of Computer and Information Science, Southwest University (e-mail: \{turou,qian05\}@email.swu.edu.cn; bhuang@swu.edu.cn).}
\thanks{Weichao Song is with the School of Software, Dalian University of Technology (e-mail: songweichao@mail.dlut.edu.cn).}
\thanks{Fangfei Ye is with the School of Traditional Chinese Medicine, Beijing University of Chinese Medicine (e-mail: 20210121138@bucm.edu.cn).}
\thanks{Haibin Ling is with the Department of Artificial Intelligence, Westlake University (e-mail: linghaibin@westlake.edu.cn).}

\thanks{$^{*}$Corresponding authors.}
}

\maketitle

\begin{abstract}
    Acupoint therapy is a core therapeutic method of Traditional Chinese Medicine (TCM), and it requires a high level of expertise and skills to detect acupoints and perform acupuncture and moxibustion.   
    Existing mixed reality (MR)-based training methods often fall short in accurate real-time detection and visualization of acupoints on the hand, limb, or torso of a real person and do not support various techniques of acupuncture and moxibustion. 
    Moreover, evaluation standards and visual guidance with fine details for each step during MR-based training are typically missing.
   To this end, we propose the MR-based TCM Acupoint Therapy Teaching System (MRATTS)---an MR-based acupoint therapy teaching and training framework. 
    MRATTS is based on a real-time hand, limb, and torso acupoint detection method to accurately track and visualize acupoints on real patients through MR. 
    On top of that, in collaboration with an experienced acupoint therapist, we design a practice method with interactive visual guidance for various acupoint therapy techniques that simulate acupressure, acupuncture (insertion, lifting-thrusting, and twisting), and moxibustion (mild, sparrow-pecking, and whirling).
    A set of TCM theory-based evaluation standards is formulated within MRATTS to enable the scoring and visualization of the accuracy and proficiency of acupoint therapy.
    The effectiveness and usefulness of MRATTS are evaluated through a controlled user study and expert feedback.
    Results of the study indicate that the MRATTS group shows clear improvements in understanding 3D locations of acupoints and proficiency in acupoint therapy compared to control groups. 
\end{abstract}


\section{Introduction}
\label{sec:MRATTS module}

\IEEEPARstart{A}{cupoint} therapy is an essential method in TCM, including acupressure, acupuncture, moxibustion, and other therapeutic techniques. Acupoints are specific points on the human body that are believed to have positive effects on health in TCM~\cite{berman2004effectiveness,Linde2016,Lee2025,Song2025,Fang2022,kim2014moxibustion,TCAAM0013}. 
All acupoint therapies revolve around these points. Therefore, learners need to master acupoint localization methods and be able to perform complex therapeutic techniques at the correct locations. 
Traditional teaching methods combine oral instruction and personal demonstration, such as the apprenticeship model~\cite{zhou2023integrating}, along with textbooks~\cite{Unschuld2010} and acupuncture bronze models~\cite{Zhu2021}. 
Some previous work improves the teaching process by using learning-based acupoint detection on 2D screens~\cite{sun2020acupoint,sun2022hand,zhang2025cad,mingluyu2025heatmap,cao2025enhanced,he2026raife,wang2025structure,zhang2024research,masood20223d,zheng2025multimodal,hu2021novel} or force feedback sensors~\cite{neto2018virtual,heng2004haptic,liao2010research}.

\begin{figure}[t]
\centering
\includegraphics[width=1.0\linewidth]{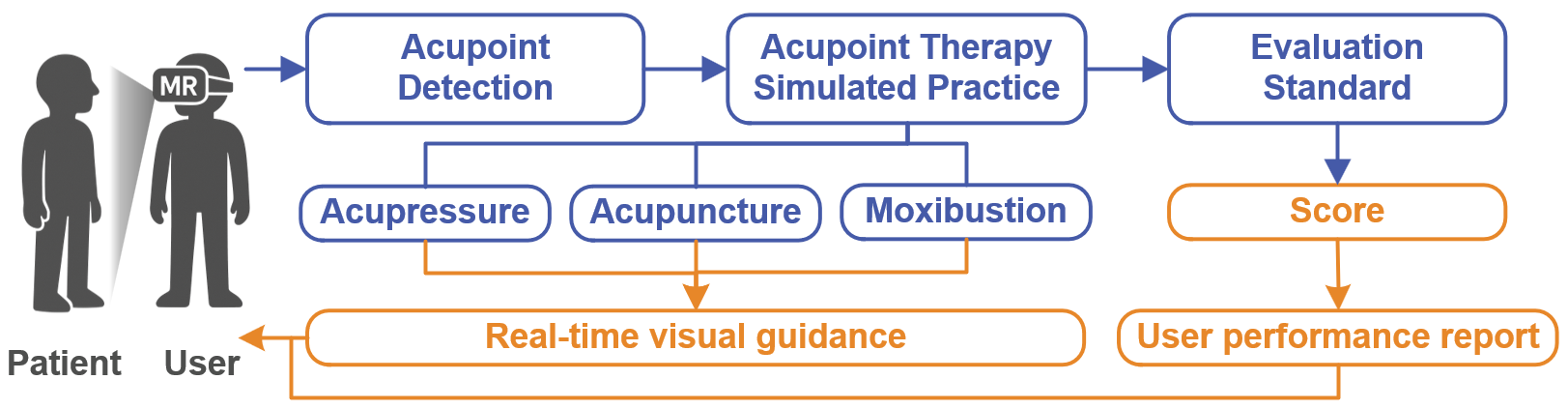}
\caption{
    MRATTS consists of three main modules (top): acupoint detection and real-time visualization on both a real person (patient) and standard human acupoint anatomical model in MR, simulated acupoint therapy practice (acupressure, acupuncture, and moxibustion) and real-time visual guidance in the HMD, and evaluation standards based on TCM theory with user performance report after simulated practice to improve user learning accuracy and efficiency. }
  \label{fig:teaser}
\end{figure}

The emergence of MR~\cite{pietschmann2023quantifying,johnson2022patient,Jadhav2023,silva2023revisiting,viglialoro2021augmented} helps users to visualize acupoints in a 3D environment and conduct immersive practices for therapeutic operations. 
Recent studies visualize facial and ear acupoints~\cite{zhang2021faceatlasar,zhang2022faceatlasar} to assist learning, and it is also possible to practice acupuncture insertion on a virtual human acupuncture anatomical model~\cite{sun2023design} similar to the acupuncture bronze model in traditional methods.

In this paper, we leverage the potentials of MR to design a full-fledged learning framework for acupoint therapies: starting from acupoint location learning, to complex therapeutic operation practice, and quantitative scoring and visualized data feedback (\Cref{fig:teaser}). 
We propose MRATTS, an MR-based acupoint therapy teaching system to address these three requirements with three main modules, as shown in~\Cref{fig:teaser} (top). 
Users can visualize acupoints on a real person through a head-mounted display (HMD), practice acupuncture and moxibustion therapies, and gain real-time visual guidance and a user performance report to enhance learning efficiency.  
We combine acupoint localization theory~\cite{wu2022interpretation, GBT12346-2021,world2008standard} in TCM with human pose estimated~\cite{lugaresi2019mediapipe} using computer vision to detect acupoint locations. 
MRATTS automatically adapts standard acupoint locations to individuals with different body types and visualizes interactive acupoint spheres overlying a real person in the HMD for subsequent practice. 
On top of this, simulated practice interactions~\cite{9089474} are designed for advanced TCM acupoint therapies, such as lifting-thrusting and twisting in acupuncture,  mild moxibustion, sparrow-pecking moxibustion, and whirling moxibustion. 
MRATTS can recognize the operations of a user and provide real-time visual guidance, e.g., color and numerical values, to assist the user with simulated practice. 
User operation data during the practice is recorded for subsequent evaluation.

In collaboration with a TCM expert, a set of full-fledged evaluation standards for the practice data is designed. 
MRATTS scores the operation of a user based on these standards and generates a performance report with data visualizations to help users improve their therapeutic techniques.

To evaluate our method, a numerical study demonstrates that the detection accuracy and runtime performance of the method meet teaching requirements. 
A subsequent user study---a controlled between-subject experiment ($N$=42, 3 groups)---verifies that MRATTS has a positive effect on teaching acupoint locations and therapeutic operations, outperforming traditional methods and a baseline previous work. 
Feedback is obtained from acupoint therapists, who note that MRATTS provides an immersive and engaging teaching method for entry-level learners with a weak foundation, improving learning efficiency.

Our contributions can be summarized as follows.
\begin{itemize}
    \item An end-to-end framework for MR-based acupoint therapy teaching.
    \item A method for MR-based hand, limb, and torso acupoint detection with real-time visualization.
    \item A simulated practice method for advanced acupoint therapies in MR, along with qualitative and quantitative evaluation standards, is proposed to assist teaching.
\end{itemize}

A preliminary, high-level overview of the MRATTS concept was previously presented as a two-page poster at the IEEE ISMAR-Adjunct conference~\cite{11236525}. This manuscript significantly extends that preliminary work by offering a comprehensive algorithmic description of the real-time acupoint localization pipeline and the detailed scoring functions of evaluation standards. Furthermore, this article introduces an in-depth system numerical evaluation and a novel six-month longitudinal evaluation to assess long-term knowledge retention, none of which were included in the initial poster presentation.

\begin{figure}[h]
  \centering
  \includegraphics[width=0.8\columnwidth]{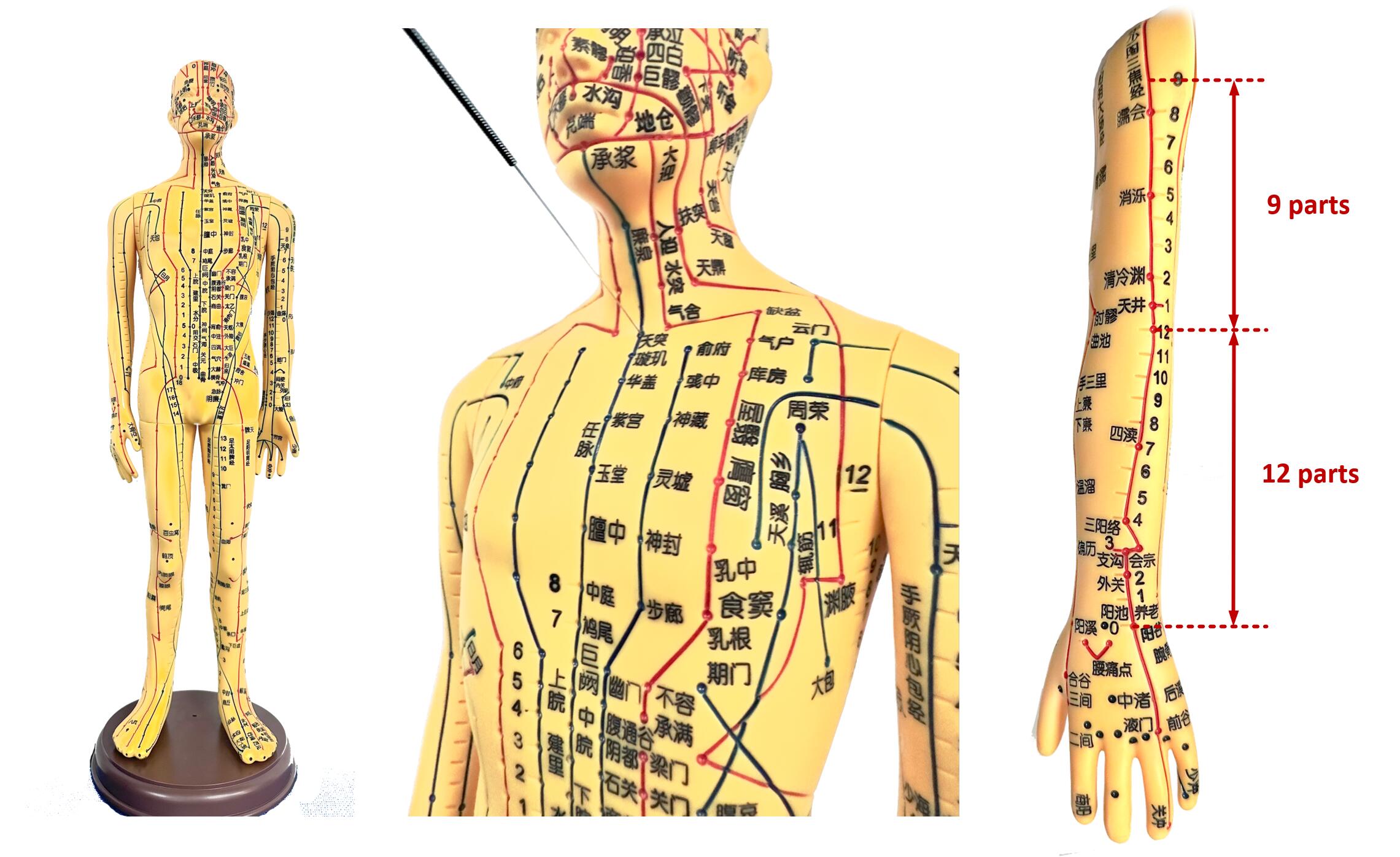}
  \caption{An acupuncture bronze model, i.e., a human acupoint model, is used for traditional acupoint location teaching.}
  \vspace{-1em}
  \label{fig:traditional acupoint loacation}
\end{figure}

\section{Medical Background}
\label{sec:Acupoint knowledge}
Accurate acupoint localization is the foundation of acupoint therapies as errors directly impact effectiveness and safety~\cite{godson2019accuracy}. 
Therefore, reliable and accurate acupoint detection is fundamental in any acupoint therapy training tool. 

Acupoint therapy consists of three primary operational techniques: acupressure, acupuncture, and moxibustion~\cite{feng2018acupoint}. In acupuncture, insertion can be divided into three types: perpendicular, oblique, and transverse insertion, based on the angle~\cite{GBT21709.20,world2008standard,Stux2003}. 
The reinforcing-reducing effects of lifting-thrusting are distinguished by variations in depth and speed~\cite{GBT21709.20,GBT21709.21}. 
For twisting, the key factors are the direction and speed of rotation~\cite{GBT21709.20,GBT21709.21}. 
Moxibustion is also categorized into three types (mild, sparrow-pecking, and whirling) depending on the motion pattern of the moxa stick~\cite{TCAAM001,GBT21709.1}. 
Each technique involves specific sets of operational parameters essential to therapeutic efficacy~\cite{lyu2019stimulation}, including spatial location, insertion angle, insertion depth, operational velocity, rotation direction, and sequence of actions as summarized in~\autoref{tab:knowledge}.
Precise control and recognition of these parameters are crucial for effective therapy and patient safety. 

Acupoint therapy manipulations are mainly categorized as reinforcing and reducing~\cite{gao2012relative}. 
Subtle variations in operational sequence and technique execution may lead to different therapeutic outcomes~\cite{huang2012transcontinental}.
For example, minor deviations in lifting-thrusting sequences or speed variations can switch a reinforcing action into a reducing effect and vice versa~\cite{GBT21709.21}. 
Therefore, an effective educational tool should allow learners to comprehend and differentiate technical details.
\begin{table}[h]
  \caption{Acupoint knowledge used in MRATTS.}
  \label{tab:knowledge}
	\centering
    \resizebox{\linewidth}{!}{
\begin{tabular}{c|c|c|c|c}
\hline
\multicolumn{2}{c|}{\textbf{Domain}}                                                                                            & \multicolumn{3}{c}{\textbf{Details}}                                                                                                                 \\ \hline
\multicolumn{2}{c|}{\multirow{5}{*}{\begin{tabular}[c]{@{}c@{}}Acupoint\\theoretical \\knowledge\end{tabular}}}   & \multicolumn{3}{c}{Acupoint name}                                                                                                                         \\ \cline{3-5} 
\multicolumn{1}{c}{}                                                                                &                               & \multicolumn{3}{c}{Acupoint location}                                                                                                                     \\ \cline{3-5} 
\multicolumn{1}{c}{}                                                                                &                               & \multicolumn{3}{c}{Acupoint indication}                                                                                                                          \\ \cline{3-5} 
\multicolumn{1}{c}{}                                                                                &                               & \multicolumn{3}{c}{Relevant organ}                                                                                                                        \\ \cline{3-5} 
\multicolumn{1}{c}{}                                                                                &                               & \multicolumn{3}{c}{Meridian}                                                                                                                       \\ \hline
\multicolumn{1}{c|}{\multirow{15}{*}{\rotatebox[origin=c]{90}{Acupoint therapeutic knowledge}}} 
& \makecell{Acu-\\pressure}                   & \multicolumn{3}{c}{Bone proportional cun (B-cun), Finger cun (F-cun)}                                                                                                         \\ \cline{2-5} 
\multicolumn{1}{c|}{}                                                                                & \multirow{11}{*}{\makecell{Acu-\\puncture}} & \multicolumn{1}{c|}{\multirow{4}{*}{\begin{tabular}[c]{@{}c@{}}Lifting and  \\ thrusting\end{tabular}}} & \multicolumn{1}{c|}{\multirow{2}{*}{\makecell{Reinforcing\\(for deficiency syndrome)}}} & First deep then shallow                 \\ \cline{5-5} 
\multicolumn{1}{c|}{}                                                                                &                               & \multicolumn{1}{c|}{}                                       & \multicolumn{1}{c|}{}                             & First fast then slow                    \\ \cline{4-5} 
\multicolumn{1}{c|}{}                                                                                &                               & \multicolumn{1}{c|}{}                                       & \multicolumn{1}{c|}{\multirow{2}{*}{\makecell{Reducing\\(for excess syndrome)}}}    & First shallow then deep                 \\ \cline{5-5} 
\multicolumn{1}{c|}{}                                                                                &                               & \multicolumn{1}{c|}{}                                       & \multicolumn{1}{c|}{}                             & First slow then fast                    \\ \cline{3-5} 
\multicolumn{1}{c|}{}                                                                                &                               & \multicolumn{1}{c|}{\multirow{4}{*}{Twisting}}              & \multicolumn{1}{c|}{\multirow{2}{*}{\makecell{Reinforcing\\(for deficiency syndrome)}}} & First clockwise then counterclockwise   \\ \cline{5-5} 
\multicolumn{1}{c|}{}                                                                                &                               & \multicolumn{1}{c|}{}                                       & \multicolumn{1}{c|}{}                             & Clockwise fast and counterclockwise slow \\ \cline{4-5} 
\multicolumn{1}{c|}{}                                                                                &                               & \multicolumn{1}{c|}{}                                       & \multicolumn{1}{c|}{\multirow{2}{*}{\makecell{Reducing\\(for excess syndrome)}}}    & First counterclockwise then clockwise   \\ \cline{5-5} 
\multicolumn{1}{c|}{}                                                                                &                               & \multicolumn{1}{c|}{}                                       & \multicolumn{1}{c|}{}                             & Counterclockwise fast and clockwise slow \\ \cline{3-5} 
\multicolumn{1}{c|}{}                                                                                &                               & \multicolumn{1}{c|}{\multirow{3}{*}{Insertion}}              & \multicolumn{2}{c}{Perpendicular insertion: 90°}                                            \\ \cline{4-5} 
\multicolumn{1}{c|}{}                                                                                &                               & \multicolumn{1}{c|}{}                                       & \multicolumn{2}{c}{Oblique insertion: 45°}                                                  \\ \cline{4-5} 
\multicolumn{1}{c|}{}                                                                                &                               & \multicolumn{1}{c|}{}                                       & \multicolumn{2}{c}{Transverse insertion: 15°}                                               \\ \cline{2-5} 
\multicolumn{1}{c|}{}                                                                                & \multirow{3}{*}{\makecell{Moxi-\\bustion}}   & \multicolumn{1}{c|}{Mild}                                   & \multicolumn{2}{c}{Position fixed 3\,$\mathrm{cm}$ from the skin}                                        \\ \cline{3-5} 
\multicolumn{1}{c|}{}                                                                                &                               & \multicolumn{1}{c|}{Sparrow-pecking}                        & \multicolumn{2}{c}{Up and down movement without fixed distance}                             \\ \cline{3-5} 
\multicolumn{1}{c|}{}                                                                                &                               & \multicolumn{1}{c|}{Whirling}                               & \multicolumn{2}{c}{Uniform reciprocating motion}                                            \\ \hline
\end{tabular}
}
\vspace{-1em}
\end{table}

\section{Related Work}

MR is used in medical education, training, and treatment~\cite{chen2017recent,viglialoro2021augmented}. Early systems for critical care training~\cite{azimi2018evaluation} and needle placement guidance~\cite{yeo2011effect,coles2011integrating} demonstrate its potential. 
While MR is applied to acupuncture training in TCM~\cite{sun2023design}, a comprehensive teaching system that integrates real-time acupoint detection, simulated therapy practice with visual guidance, and empirical evaluation standards remains understudied.

\vspace{-1em}
\subsection{Acupoint Detection and Visualization}
Diagrams and physical models such as the acupuncture bronze model shown in \Cref{fig:traditional acupoint loacation} are traditional approaches for localizing acupoints. 
To improve acupoint visualization, 3D anatomical models are introduced to integrate skin and organs~\cite{cao2015implementation}, vascular and muscular systems are added~\cite{neto2018virtual}, and meridians are included for immersive MR learning~\cite{sun2023design}. 
However, these static models do not adapt to the diverse body types of patients.

AI-driven methods are available to automate acupoint detection, although these are often limited to 2D screens and lack real-time performance. 
For instance, deep Convolutional Neural Networks (CNNs) are used for 2D arm acupoint detection~\cite{sun2020acupoint}, while an improved HRNet is used for hand acupoints~\cite{sun2022hand,zhang2025cad,mingluyu2025heatmap}. To enhance real-time speed and robustness, YOLOv11~\cite{cao2025enhanced} and RAIFE-Net~\cite{he2026raife} are further introduced. Furthermore, TCM bone-proportional measurement theory is integrated to refine acupoint localization~\cite{wang2025structure,zhang2024research}.

Similarly, 3D locations of acupoints are detected using RGB-D data for hands~\cite{masood20223d} or fused thermal-depth imaging for the torso~\cite{zheng2025multimodal}. 
While real-time visualization for torso and limb acupoints is achieved~\cite{hu2021novel}, all these visualizations are displayed on 2D screens. 
AR offers a more immersive solution:  FaceatlasAR is available for real-time detection on the face~\cite{zhang2021faceatlasar} and later the ears~\cite{zhang2022faceatlasar}. 
However, existing HMD-based works~\cite{butaslac2022systematic,zhang2022faceatlasar,chen20213d} are predominantly AR-focused, without full-body visualization, and lack advanced interaction design beyond simple visual overlays.

\begin{table*}[tb]
  \caption{A comparison of different acupoint training methods.}
  \label{tab:method comparison}
  \centering
  \resizebox{\linewidth}{!}{
\begin{tabular}{l|c|c|c|c|c|c|c|ccc|c|c}
\hline
    &  & \multicolumn{4}{c|}{\textbf{Real patient Acupoint Visualization}} &  &  & \multicolumn{3}{c|}{\textbf{Acupuncture Visualization}} &  \\ \cline{3-6} \cline{9-11}
\multirow{-2}{*}{\textbf{Representative Methods}} & \multirow{-2}{*}{\textbf{MR}} & \textbf{Hand} & \textbf{Facial} & \textbf{Limb} & \textbf{Torso} & \multirow{-2}{*}{\begin{tabular}[c]{@{}c@{}}\textbf{Real-time}\\ \textbf{Visualization}\end{tabular}} & \multirow{-2}{*}{\begin{tabular}[c]{@{}c@{}}\textbf{Has Acupoint}\\ \textbf{Model}\end{tabular}} & \multicolumn{1}{c|}{\textbf{Insertion}} & \multicolumn{1}{c|}{\textbf{Lifting-Thrusting}} & \textbf{Twisting} & \multirow{-2}{*}{\textbf{Moxibustion}} & \multirow{-2}{*}{\begin{tabular}[c]{@{}c@{}}\textbf{Evaluation}\\ \textbf{Standards}\end{tabular}} \\ \hline

Acu. Training System~\cite{sun2023design}& \cellcolor[HTML]{80D980}\ding{51} & \cellcolor[HTML]{CCCCCC}$\bm{\times}$ & \cellcolor[HTML]{CCCCCC}$\bm{\times}$ & \cellcolor[HTML]{CCCCCC}$\bm{\times}$ & \cellcolor[HTML]{CCCCCC}$\bm{\times}$ & \cellcolor[HTML]{CCCCCC}$\bm{\times}$ & \cellcolor[HTML]{80D980}$\checkmark$ & \multicolumn{1}{c|}{\cellcolor[HTML]{80D980}$\checkmark$} & \multicolumn{1}{c|}{\cellcolor[HTML]{CCCCCC}$\bm{\times}$} & \cellcolor[HTML]{CCCCCC}$\bm{\times}$ & \cellcolor[HTML]{CCCCCC}$\bm{\times}$ & \cellcolor[HTML]{FFFF00}\LEFTcircle \\ \hline

E-faceatlasAR~\cite{zhang2022faceatlasar}& \cellcolor[HTML]{80D980}\ding{51} & \cellcolor[HTML]{CCCCCC}$\bm{\times}$ & \cellcolor[HTML]{80D980}$\checkmark$ & \cellcolor[HTML]{CCCCCC}$\bm{\times}$ & \cellcolor[HTML]{CCCCCC}$\bm{\times}$ & \cellcolor[HTML]{80D980}$\checkmark$ & \cellcolor[HTML]{CCCCCC}$\bm{\times}$ & \multicolumn{1}{c|}{\cellcolor[HTML]{CCCCCC}$\bm{\times}$} & \multicolumn{1}{c|}{\cellcolor[HTML]{CCCCCC}$\bm{\times}$} & \cellcolor[HTML]{CCCCCC}$\bm{\times}$ & \cellcolor[HTML]{CCCCCC}$\bm{\times}$ & \cellcolor[HTML]{CCCCCC}\Circle \\ \hline

MRUCT~\cite{wang2025mruct} & \cellcolor[HTML]{80D980}\ding{51} & \cellcolor[HTML]{FFFF00}\textbf{--} & \cellcolor[HTML]{FFFF00}\textbf{--} & \cellcolor[HTML]{80D980}$\checkmark$ & \cellcolor[HTML]{FFFF00}\textbf{--} & \cellcolor[HTML]{CCCCCC}$\bm{\times}$ & \cellcolor[HTML]{CCCCCC}$\bm{\times}$ & \multicolumn{1}{c|}{\cellcolor[HTML]{80D980}$\checkmark$} & \multicolumn{1}{c|}{\cellcolor[HTML]{CCCCCC}$\bm{\times}$} & \cellcolor[HTML]{CCCCCC}$\bm{\times}$ & \cellcolor[HTML]{CCCCCC}$\bm{\times}$ & \cellcolor[HTML]{CCCCCC}\Circle \\ \hline

Hap-AcuMR~\cite{guruge2025advancing} & \cellcolor[HTML]{80D980}\ding{51} & \cellcolor[HTML]{CCCCCC}$\bm{\times}$ & \cellcolor[HTML]{CCCCCC}$\bm{\times}$ & \cellcolor[HTML]{CCCCCC}$\bm{\times}$ & \cellcolor[HTML]{CCCCCC}$\bm{\times}$ & \cellcolor[HTML]{CCCCCC}$\bm{\times}$ & \cellcolor[HTML]{80D980}$\checkmark$ & \multicolumn{1}{c|}{\cellcolor[HTML]{80D980}$\checkmark$} & \multicolumn{1}{c|}{\cellcolor[HTML]{CCCCCC}$\bm{\times}$} & \cellcolor[HTML]{CCCCCC}$\bm{\times}$ & \cellcolor[HTML]{CCCCCC}$\bm{\times}$ & \cellcolor[HTML]{FFFF00}\LEFTcircle \\ \hline

TCM Massage Robot~\cite{hu2021novel}& \cellcolor[HTML]{fc8d59}\ding{55} & \cellcolor[HTML]{CCCCCC}$\bm{\times}$ & \cellcolor[HTML]{CCCCCC}$\bm{\times}$ & \cellcolor[HTML]{80D980}$\checkmark$ & \cellcolor[HTML]{80D980}$\checkmark$ & \cellcolor[HTML]{80D980}$\checkmark$ & \cellcolor[HTML]{CCCCCC}$\bm{\times}$ & \multicolumn{1}{c|}{\cellcolor[HTML]{CCCCCC}$\bm{\times}$} & \multicolumn{1}{c|}{\cellcolor[HTML]{CCCCCC}$\bm{\times}$} & \cellcolor[HTML]{CCCCCC}$\bm{\times}$ & \cellcolor[HTML]{CCCCCC}$\bm{\times}$ &  \cellcolor[HTML]{CCCCCC}\Circle \\ \hline

RGB-D CNN~\cite{masood20223d}& \cellcolor[HTML]{fc8d59}\ding{55} & \cellcolor[HTML]{80D980}$\checkmark$ & \cellcolor[HTML]{CCCCCC}$\bm{\times}$ & \cellcolor[HTML]{CCCCCC}$\bm{\times}$ & \cellcolor[HTML]{CCCCCC}$\bm{\times}$ & \cellcolor[HTML]{FFFF00}$\bm{\sim}$ & \cellcolor[HTML]{CCCCCC}$\bm{\times}$ & \multicolumn{1}{c|}{\cellcolor[HTML]{CCCCCC}$\bm{\times}$} & \multicolumn{1}{c|}{\cellcolor[HTML]{CCCCCC}$\bm{\times}$} & \cellcolor[HTML]{CCCCCC}$\bm{\times}$ & \cellcolor[HTML]{CCCCCC}$\bm{\times}$ & \cellcolor[HTML]{CCCCCC}\Circle \\ \hline

Deep CNN~\cite{sun2020acupoint} & \cellcolor[HTML]{fc8d59}\ding{55} & \cellcolor[HTML]{CCCCCC}$\bm{\times}$ & \cellcolor[HTML]{CCCCCC}$\bm{\times}$ & \cellcolor[HTML]{80D980}$\checkmark$ & \cellcolor[HTML]{CCCCCC}$\bm{\times}$ & \cellcolor[HTML]{FFFF00}$\bm{\sim}$ & \cellcolor[HTML]{CCCCCC}$\bm{\times}$ & \multicolumn{1}{c|}{\cellcolor[HTML]{CCCCCC}$\bm{\times}$} & \multicolumn{1}{c|}{\cellcolor[HTML]{CCCCCC}$\bm{\times}$} & \cellcolor[HTML]{CCCCCC}$\bm{\times}$ & \cellcolor[HTML]{CCCCCC}$\bm{\times}$ &  \cellcolor[HTML]{CCCCCC}\Circle \\ \hline

Improved HRNet~\cite{sun2022hand,zhang2025cad,mingluyu2025heatmap}& \cellcolor[HTML]{fc8d59}\ding{55} & \cellcolor[HTML]{80D980}$\checkmark$ & \cellcolor[HTML]{CCCCCC}$\bm{\times}$ & \cellcolor[HTML]{CCCCCC}$\bm{\times}$ & \cellcolor[HTML]{CCCCCC}$\bm{\times}$ & \cellcolor[HTML]{CCCCCC}$\bm{\times}$ & \cellcolor[HTML]{CCCCCC}$\bm{\times}$ & \multicolumn{1}{c|}{\cellcolor[HTML]{CCCCCC}$\bm{\times}$} & \multicolumn{1}{c|}{\cellcolor[HTML]{CCCCCC}$\bm{\times}$} & \cellcolor[HTML]{CCCCCC}$\bm{\times}$ & \cellcolor[HTML]{CCCCCC}$\bm{\times}$ & \cellcolor[HTML]{CCCCCC}\Circle \\ \hline

Ours & \cellcolor[HTML]{80D980}\ding{51} & \cellcolor[HTML]{80D980}$\checkmark$ & \cellcolor[HTML]{CCCCCC}$\bm{\times}$ & \cellcolor[HTML]{80D980}$\checkmark$ & \cellcolor[HTML]{80D980}$\checkmark$ & \cellcolor[HTML]{80D980}$\checkmark$ & \cellcolor[HTML]{80D980}$\checkmark$ & \multicolumn{1}{c|}{\cellcolor[HTML]{80D980}$\checkmark$} & \multicolumn{1}{c|}{\cellcolor[HTML]{80D980}$\checkmark$} & \cellcolor[HTML]{80D980}$\checkmark$ & \cellcolor[HTML]{80D980}$\checkmark$ & \cellcolor[HTML]{80D980}\CIRCLE \\ \hline
\end{tabular}
  }
  
  \vspace{3pt}
  \raggedright \scriptsize
  * \textbf{--}: Not mentioned in the paper; $\bm{\sim}$: Slower speed; \LEFTcircle: Partially included.
\end{table*}

\vspace{-1em}

\subsection{Simulated Acupoint Therapy Practice}

Works in simulated acupoint therapy primarily focus on acupuncture. 
Early works include replication of acupuncture in virtual patient environments~\cite{heng2004haptic}, simulation of skin deformation during needle insertion~\cite{liao2010research}, and integration of haptic feedback with AR~\cite{coles2011integrating}. 
More recent methods involve automatic hand tracking with combined visual and tactile feedback~\cite{9865994}, and interactive MR simulations with numerical data on needle parameters~\cite{sun2023design}. 
However, these approaches often lack full immersion or are based on generic models rather than real patient scenarios. While most simulation systems rely on virtual models, MRUCT demonstrates the possibility of high-precision needle guidance on real human subjects~\cite{wang2025mruct}.

To the best of our knowledge, there is no simulation practice for acupressure and moxibustion in MR. 
Existing acupuncture simulations are also limited, rarely incorporating advanced techniques, e.g., lifting-thrusting, twisting, or foundational TCM principles like reinforcing-reducing theory. 
Furthermore, most rely on virtual models rather than practice on real human subjects within an MR environment, see ~\Cref{sec:discussion} for a discussion.

\vspace{-1em}
\subsection{Acupoint Therapy Evaluation}
Utilizing sensory feedback to aid learners in self-correcting behavior enhances learning outcomes. 
Visual feedback is utilized for guiding spatial movements~\cite{7105411}.
In acupuncture training, color-coded cues are used for correct locations~\cite{9865994}, and task panels showing numerical data such as depth and angle are used to guide practice~\cite{sun2023design}. Recent work incorporates wearable haptic devices for insertion depth-responsive tactile guidance~\cite{guruge2025advancing}.

However, such feedback can be too simplistic to help learners refine their techniques.
Moreover, evaluations of practice in these works are often binary, i.e., correct/incorrect, and comprehensive performance scores are not available after a practice session.
An expert-weighted algorithm covering depth, angle, and duration helps distinguish user proficiency. However, these methods are limited to simple needle insertion. They lack coverage of complex therapeutic operations, such as lifting-thrusting, twisting, and moxibustion, as categorized in \cite{guruge2025advancing}.

\vspace{-1em}
\subsection{Comparison with Similar Work} 
The work of Sun et al.~\cite{sun2023design} is close to ours, but differs in several important aspects: (a) our real-time acupoint detection method enables users to learn acupoint localization on different patients in real-time, whereas a fixed model is used in previous work.
(b) Advanced acupuncture and moxibustion techniques needed in real treatments can be learned with our method, while only simple needle insertions are included in previous work.
(c) Our evaluation standards provide both qualitative and quantitative scores, along with data visualization through charts, to help users correct operational errors.

In addition to the recent work of Sun et al.~\cite{sun2023design}, we also compare our method to other acupoint training methods as shown in \autoref{tab:method comparison}.
It can be seen that existing methods focus on acupoint therapy specific to certain regions, e.g., hands, face, limbs, or torso, and they typically do not support the practice of techniques other than insertion. 
Moreover, moxibustion is not covered by these works, and most works do not provide evaluations of user performance.

In comparison, our new method is more comprehensive in body regions coverage, therapy techniques supported, and a full set of evaluation standards is provided as shown in the last line of~\autoref{tab:method comparison}.

\begin{figure*}[!htb]
  \centering 
  \includegraphics[width=1.0\linewidth]{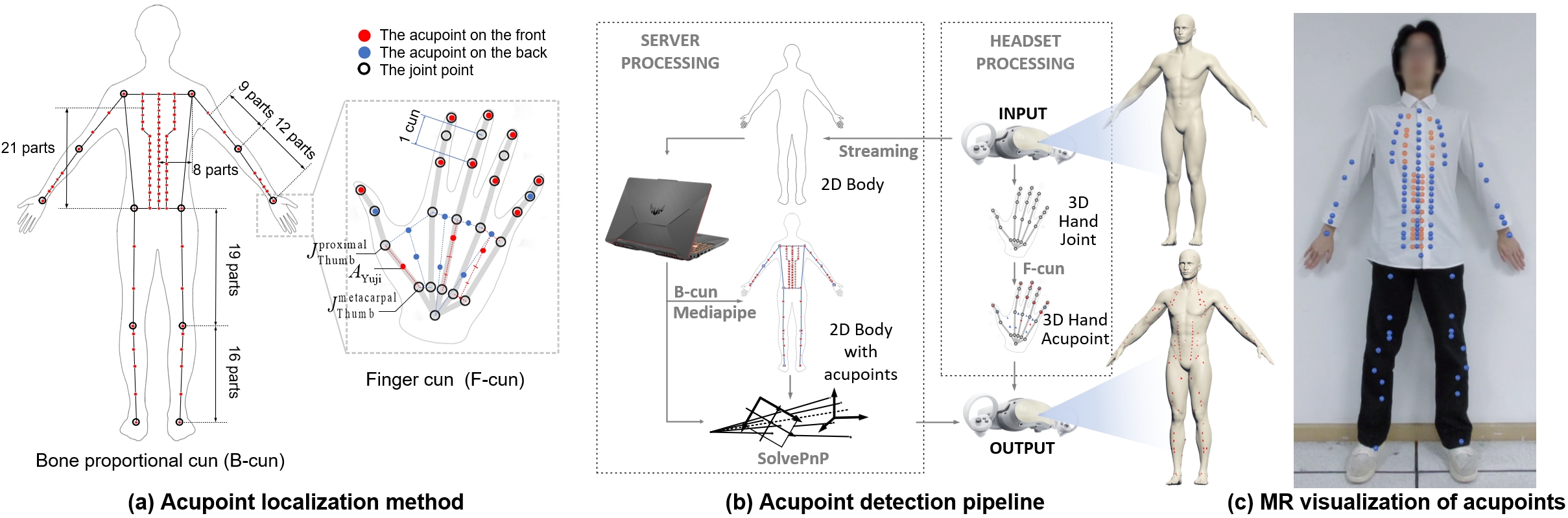}
  \caption{
We use (a) the bone proportional cun (B-cun) and finger cun (F-cun) methods~\cite{world2008standard,GBT12346-2021} for locating acupoints. The joint points are detected with a computer vision-based technique (b). The detected acupoints are visualized as (c) colored spheres overlaid on the real person shown in the HMD.  
  }
  \vspace{-1em}
  \label{fig:acupoint detection and visualization}
\end{figure*}

\section{Methods}

In collaboration with an acupoint expert (E1) who has undergone over four years of professional training, we identified three core Design Requirements (\textbf{DR1–DR3}) for a comprehensive acupoint therapy teaching system. These requirements guide the design of MRATTS, consisting of three modules detailed in the following sections.

\textbf{DR1: Real-time and Adaptive Acupoint Localization.} Users need to learn accurate acupoint detection on diverse body types rather than static models.  Our Acupoint Detection and Visualization module (\Cref{sec:Acupoint detection and visualization}) addresses this requirement.

\textbf{DR2: Authentic Simulation of Advanced Techniques.} Teaching must go beyond simple needle insertion to include complex manipulations like lifting-thrusting and various moxibustion patterns. Therefore, the Simulated Practice and Visual Guidance module detailed in ~\Cref{sec:Simulated practice and visual guidance} is devised.

\textbf{DR3: Quantitative and Interpretable Feedback.} A robust evaluation standard is required to provide feedback more than binary ``correct/incorrect". The tool must offer detailed scoring based on TCM theory, specifically incorporating treatment effectiveness (to ensure therapeutic efficacy) and risk level (to prevent operational injuries, such as excessive needle depth or skin burning). This is addressed by our Evaluation Standards and User Performance Report module as shown in ~\Cref{sec:Evaluation standards}.

The entire learning process is evaluated, with corrections suggested using the user performance report.
By integrating these modules, MRATTS provides an end-to-end instructional workflow for learners. Specifically, users first visualize and learn to localize acupoints directly on a real person to account for individual anatomical variations (addressing \textbf{DR1}). They then proceed to practice advanced therapeutic techniques, such as acupuncture and moxibustion, with real-time visual assistance and guidance (addressing \textbf{DR2}). Subsequently, the tool evaluates their performance based on TCM-informed scoring functions, generating a comprehensive report with data visualizations and scores to facilitate self-correction and skill refinement (addressing \textbf{DR3}).

\subsection{Acupoint Detection and Visualization}
\label{sec:Acupoint detection and visualization}

Acupoint detection and visualization should consider different body shapes. Therefore, it is challenging to learn effectively solely through the standardized Acupuncture Bronze Model (\Cref{fig:traditional acupoint loacation}).

\begin{CJK*}{UTF8}{gbsn}
Two types of acupoint detection methods exist in the TCM theory: the anatomical landmark method and the proportional measurement method. 
The anatomical landmark method~\cite{GBT12346-2021,world2008standard} determines acupoint locations based on fixed surface landmarks. 
However, it is applicable to only a limited number of acupoints. 
In contrast, the proportional measurement method can cover most of the acupoints by using the Bone proportional cun (B-cun) and Finger cun (F-cun), where ``\emph{cun}''---寸\quad indicates a standard relative unit of measurement used to locate acupoints based on the proportions of the individual's body. 
B-cun and F-cun ensure localization precision by accounting for variations in body size and shape among different individuals~\cite{GBT12346-2021,world2008standard}. 
\end{CJK*}

Therefore, our goal is to devise an acupoint localization algorithm using the proportional measurement that uses joints as primary landmarks to calculate the length and width of various body parts as a proportional basis for acupoint localization~\cite{GBT12346-2021}. 
As shown in \Cref{fig:acupoint detection and visualization}(a), the positions of all acupoints (indicated by red and blue circles) are derived proportionally from the skeletal joints (black rings). 
For instance, TCM defines the location of the Yuji (LU10) acupoint as the ``midpoint of the radial side of the first metacarpal bone", as shown on the right side of \Cref{fig:acupoint detection and visualization}(a):  
\begin{equation}
    \boldsymbol{A}_{\text{Yuji}} = \frac{\boldsymbol{J}_{\text{Thumb}}^{\text{proximal}} + \boldsymbol{J}_{\text{Thumb}}^{\text{metacarpal}}}{2}\;\nonumber.
    \label{eq:yuji_calculation}
\end{equation}

In the generalized B/F-cun calculation, the target acupoint  $\boldsymbol{A}_\text{target} \in \mathbb{R}^3$ is located  between two reference points $\boldsymbol{P}_1$ and $\boldsymbol{P}_2$ with a ratio $\lambda$, as shown in \Cref{eq:generalized_bcun}:  
\begin{equation}
   \boldsymbol{A}_\text{target} = f(\boldsymbol{P}_1 ,\boldsymbol{P}_2) = (1 - \lambda)\boldsymbol{P}_1 + \lambda\boldsymbol{P}_2
    \label{eq:generalized_bcun}\;.
\end{equation}

A reference point $\boldsymbol{P} \in \mathbb{R}^3$ is defined such that 
$\boldsymbol{P} \in \mathcal{J} \cup \{f(\boldsymbol{J}_i, \boldsymbol{J}_j)\}$, 
indicating it can be either a physical joint $\boldsymbol{J} \in \mathcal{J}$ 
or a virtual point derived from a set of joints via the linear interpolation function $f(\cdot)$. All potential $\boldsymbol{A}_\text{target}$ localized via this method form the set of detectable acupoints $\mathcal{A}_{\text{detected}}$. Therefore, it is important to automatically and robustly detect these reference joint points from the video stream of the HMD in our MR method. 

The workflow of our acupoint detection pipeline is illustrated in \Cref{fig:acupoint detection and visualization}(b).
We utilize the power of a CNN to detect the skeleton of a person from the video stream,  apply the proportional measurement method on the skeleton thereof, and then transform the locations back to the 3D space of the HMD.
The CNN is trained using spatial structural priors of human anatomy,
and the process of extracting 3D joint positions from a 2D video stream $\boldsymbol{X}$ can be formulated as a regression problem:
\begin{equation}
    \hat{\boldsymbol{Y}} = \varPhi(\boldsymbol{X}) = \{ (x_k, y_k, z_k, v_k) \mid k = 1, \dots, K \}\;,
    \label{eq:pose_regression}
\end{equation}
where $\Phi$ denotes the non-linear mapping function of the CNN-based architecture.
Specifically, the lightweight backbone used in MediaPipe~\cite{lugaresi2019mediapipe}, which regresses the input image to skeletal coordinates, is adopted in this work.

However, the raw output $\hat{\boldsymbol{Y}}$ lacks the absolute depth and spatial alignment required for the HMD's ego-centric view.
To transform the raw prediction $\hat{\boldsymbol{Y}}$ to the camera space of the HMD, we apply the Perspective-n-Point algorithm. 
Using the intrinsic matrix $\mathbf{K}$ derived from the FOV, we estimate the skeletal pose and depth $Z_{cam}$. 
Next, we back-project the 2D coordinates $[u, v]$ using $\mathbf{K}^{-1}$ and $Z_\text{cam}$ to obtain precise 3D positions $\boldsymbol{P}_\text{cam}$ aligned with the user's view:
\begin{equation}
    \boldsymbol{P}_\text{cam} = Z_\text{cam} \cdot \mathbf{K}^{-1} \cdot \begin{bmatrix} u \\ v \\ 1 \end{bmatrix}\;.
    \label{eq:back_projection}
\end{equation}

\subsection{Simulated Practice and Visual Guidance}
\label{sec:Simulated practice and visual guidance}
The detected acupoints $\mathcal{A}_{\text{detected}}$ are the basis for simulated practices in our work, including acupressure, acupuncture, and moxibustion.
These practices are aided by associated visual guidance.
\subsubsection{Acupressure}

Acupressure involves applying finger pressure to specific acupoints on the body to promote health benefits. 
While the thumb is traditionally preferred for its greater force application, we employ the index finger to facilitate the learning of precise acupoint localization and align with established interaction paradigms in MR. 

\subsubsection{Acupuncture}
\label{sec:acupuncture}
In acupuncture practices, the physician manually controls the movement of a needle throughout the entire process, from the initial pickup to the insertion into the acupoint. 
Following insertion, the physician typically releases the grip to retain the needle at a specific acupoint for a designated duration, aiming to achieve the therapeutic goals of TCM. 
Moreover, acupuncture is more than merely inserting a needle into a specific location. 
It also includes a series of complex manipulation techniques for treatment, namely, insertion, lifting-thrusting, and twisting.

To authentically simulate this procedure, we employ bare-hand interaction, utilizing a pinch gesture to simulate the authentic gripping mechanics of holding a needle~\cite{Yang2019, Monteiro2021,lee2010hand,wang2023comparative}. 
We use a dedicated collision trigger~\cite{xie2012research} to detect insertion events by monitoring the contact between the needle tip and the target acupoint sphere. 
We carefully simulate the three important techniques in MRATTS to provide a detailed procedure learning experience.
\begin{figure}[htb]
  \centering 
  \includegraphics[width=\linewidth]{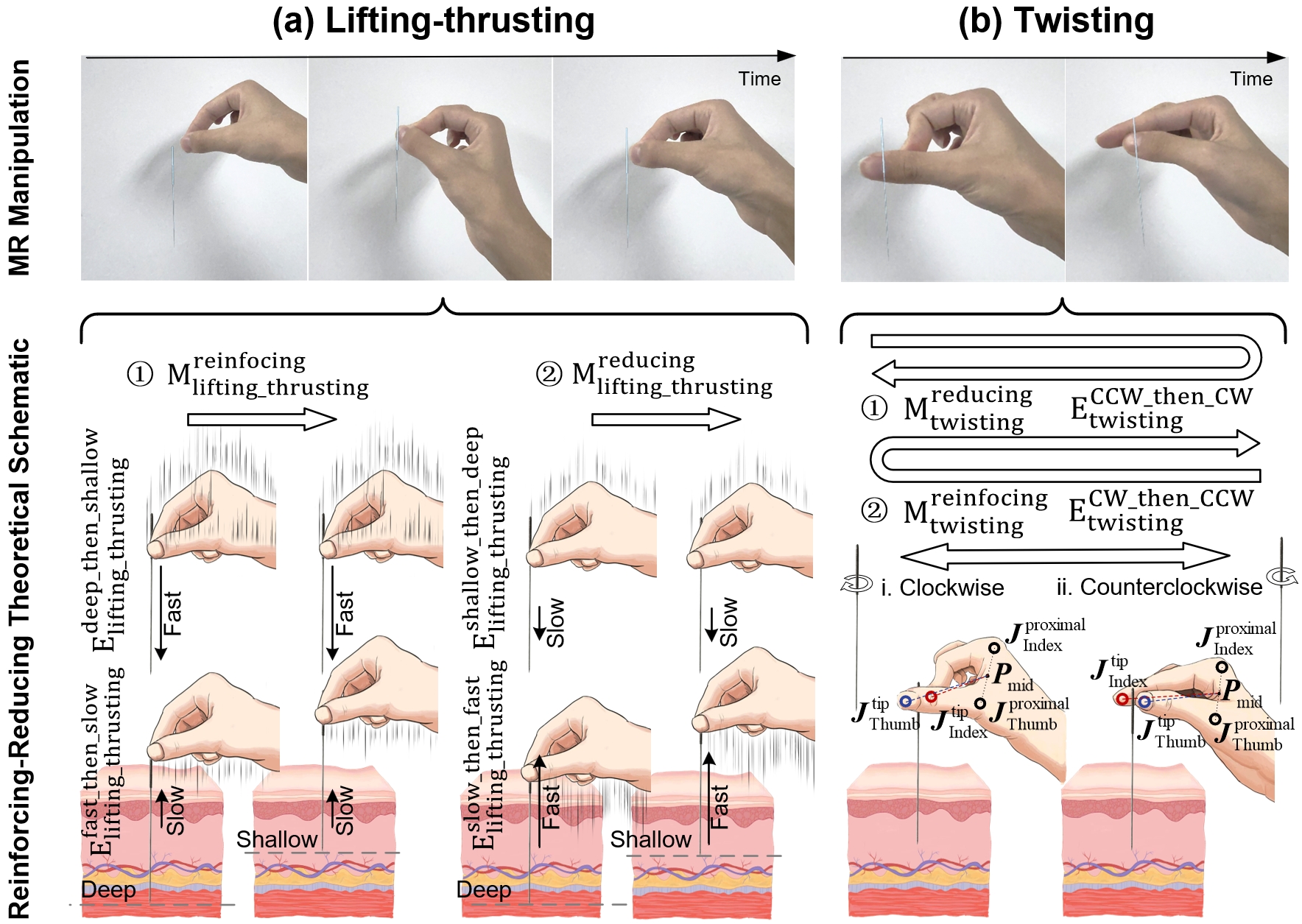}
  \caption{Simulated practice of (a) lifting-thrusting and (b) twisting in acupuncture in MRATTS. Manipulation sequences in MR are shown on top, while the simulated patterns of reinforcing and reducing techniques are illustrated at the bottom.  }
      \label{fig:Acupuncture}
\end{figure}

\textbf{Insertion:} insertion techniques are categorized into perpendicular, oblique, and transverse insertion, distinguished by the angle of entry. 
We determine the insertion type of the user by calculating the angle between the needle's direction vector and the skin's surface normal at the insertion point at the moment the trigger event occurs.

\textbf{Lifting-thrusting:} 
lifting-thrusting operations are classified into reinforcing and reducing methods (\Cref{fig:Acupuncture}(a)). 
The reinforcing method $\mathrm{M}_{\mathrm{lift\_thrust}}^{\mathrm{reinforce}}$---characterized by ``\textit{fast thrusting with slow lifting $\mathrm{E}_{\mathrm{lift\_thrust}}^{\mathrm{fast\_then\_slow}}$}" and ``\textit{deep needling followed by shallow needling $\mathrm{E}_{\mathrm{lift\_thrust}}^{\mathrm{deep\_then\_shallow}}$}"---is illustrated in \Cref{fig:Acupuncture}(a--bottom left). 
Conversely, the opposite operation shown in \Cref{fig:Acupuncture}(a--bottom right) indicates the reducing method.

Simulating the lifting-thrusting technique requires precise identification of needle depth and speed to distinguish between the two manipulation types. 
Following the detection of the insertion event, we continuously update the maximum needle penetration length and record the corresponding timestamp. 
The actual vertical insertion depth is then calculated by combining this length with the insertion angle~\cite{sun2023design}. 
Furthermore, by recording the timestamps for the initial insertion (start), the maximum depth point, and the final withdrawal (end), we can determine the duration of both the thrusting (downward) and lifting (upward) phases. 
Therefore, the average speed for each phase can be derived.

Identifying the chronological sequence of depth changes, i.e., deep needling followed by shallow needling, requires two consecutive cycles. Therefore, we define a single practice session with two lifting-thrusting operations, serving as the minimum distinguishable unit to differentiate between reinforcing and reducing techniques.  
With this design, MRATTS can accurately classify the specific type of manipulation.

\textbf{Twisting:} twisting is also categorized into reinforcing versus reducing,  distinguished by the chronological sequence of clockwise and counter-clockwise rotations (\Cref{fig:Acupuncture}(b)). 
The twisting reinforcing method $\mathrm{M}_{\mathrm{twist}}^{\mathrm{reinforce}}$ is defined as a composite movement of ``\textit{first clockwise, followed by counter-clockwise} $\mathrm{E}_{\mathrm{twist}}^{\mathrm{CW\_then\_CCW}}$".
Conversely, the twisting reducing method follows the reverse sequence.
To automatically classify the twisting type, the critical step involves detecting the occurrence of transitions between state \circled{1} and state \circled{2} in \Cref{fig:Acupuncture}(b).
We observe that in state \circled{1}, the distance from the thumb tip to the midpoint of the thumb and index  finger roots is significantly greater than that of the index fingertip, whereas the opposite is observed in state \circled{2}. 

Therefore, by calculating the distance difference
\begin{equation}
    \Delta d = \| \boldsymbol{P}_{\text{mid}} - \boldsymbol{J}_{\text{Thumb}}^{\text{tip}} \| - \| \boldsymbol{P}_{\text{mid}} - \boldsymbol{J}_{\text{Index}}^{\text{tip}} \|\;,
    \label{eq:twisting joint}
\end{equation}
where $\boldsymbol{J}_{\text{Thumb}}^{\text{tip}}$ and $\boldsymbol{J}_{\text{Index}}^{\text{tip}}$ represent the thumb and index fingertips respectively, and $\boldsymbol{P}_{\text{mid}}$ denotes the midpoint between the index finger root $\boldsymbol{J}_{\text{Index}}^{\text{proximal}}$ and the thumb finger root $\boldsymbol{J}_{\text{Thumb}}^{\text{proximal}}$, we can determine whether the current gesture is in state \circled{1} ($\Delta d > 0$) or state \circled{2} ($\Delta d < 0$) based on the sign of $\Delta d$. Consequently, monitoring the sequence of sign changes in $\Delta d$ allows for the logical determination of whether the twisting action is i. clockwise or ii. counter-clockwise.
In the twisting simulation practice, users are also required to complete  at least  two consecutive cycles.

In addition to the simulation of different techniques, we provide users with multiple needles of different sizes~\cite{lee2018quantitative,ellis1991fundamentals} for selection, enabling the learning for choosing the appropriate type of needle for the corresponding sites of the body.

\begin{figure}[htb]
  \centering
  \includegraphics[width=\linewidth]{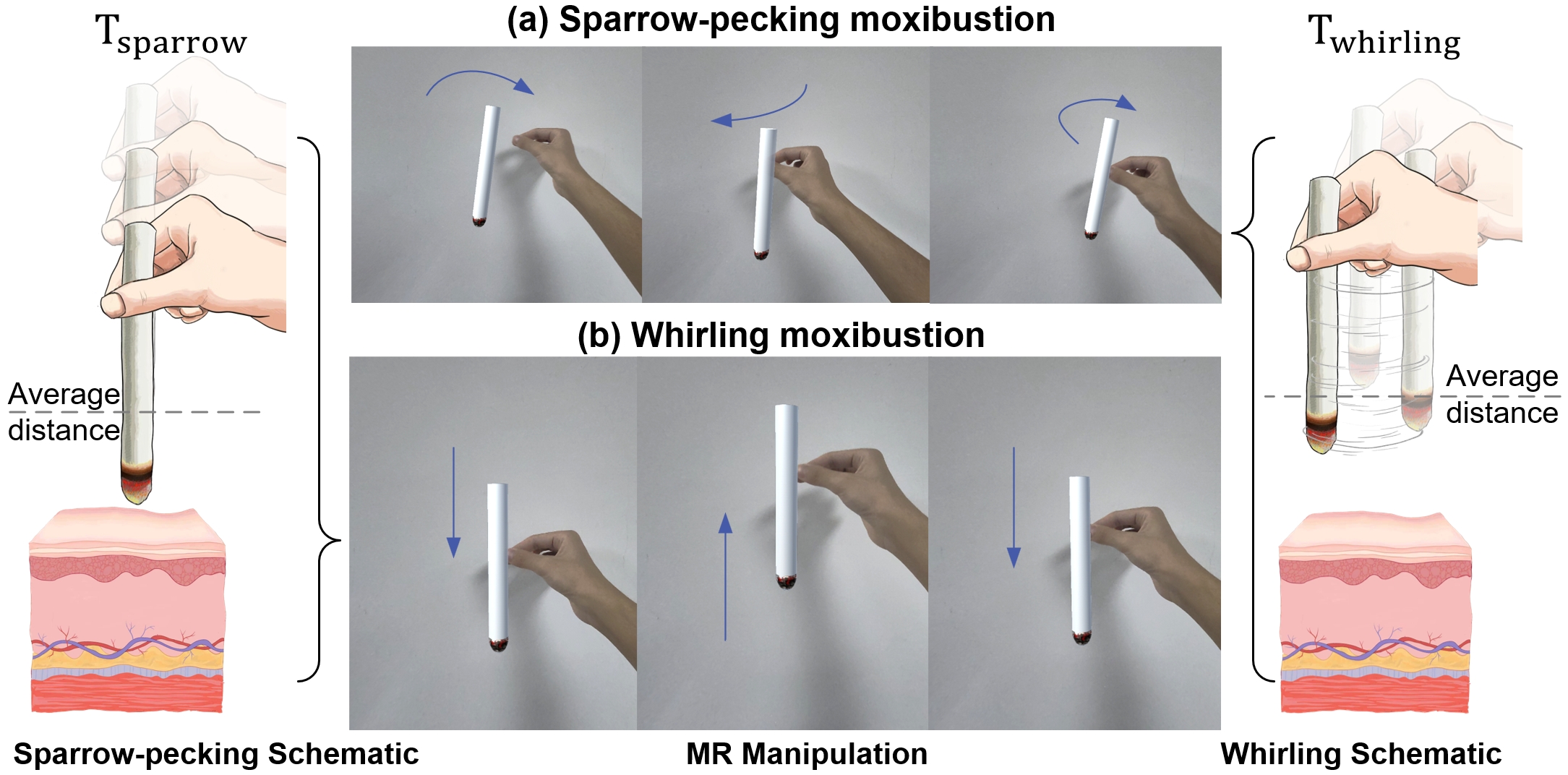}
  \caption{ MRATTS supports the practice of two major moxibustion techniques: (a) sparrow-pecking, and (b) whirling. Interactions in MR shown in the middle simulate the illustrated operations on the left and right.    }
  \label{fig:Moxibustion}
\end{figure}

\subsubsection{Moxibustion}   
Moxibustion is a therapy that utilizes the heat generated by burning moxa to induce therapeutic effects on the human body. 
Three main techniques, namely, mild, sparrow-pecking, and whirling moxibustion, are simulated in MRATTS (\Cref{fig:Moxibustion}). 
Mild moxibustion is the most straightforward one, characterized by holding the moxa stick static above the target acupoint $\boldsymbol{A}_\text{target}$. 
Sparrow-pecking moxibustion involves a vertical up-and-down movement over the acupoint, mimicking a bird pecking at food, as illustrated in \Cref{fig:Moxibustion}(a). 
Whirling moxibustion entails moving the stick in a repetitive circular motion over the acupoint, as shown in \Cref{fig:Moxibustion}(b).

We identify the moxibustion technique based on the motion data of the moxa stick's head. 
Mild moxibustion can be identified by recording the maximum and minimum distances between the moxa stick head and the target acupoint during the practice session.  
This is based on the assumption that the position of the moxa stick remains nearly stationary under correct execution. 
The main distinction between sparrow-pecking and whirling moxibustion lies in their motion patterns.  Sparrow-pecking maintains a trajectory almost strictly aligned with the vertical surface normal of the skin at the acupoint.  In contrast, whirling moxibustion deviates from this axis, creating a noticeable angle with the vertical normal vector. 
Therefore, we quantify the number of instances (time frames) during the operation where the angle exceeds a specific threshold, and the movement possesses a certain speed.  
A higher count indicates a higher probability of rotational motion (whirling moxibustion).  
Simultaneously, incorporating speed as one of the criteria prevents static mild moxibustion performed in an incorrect position from being misclassified as whirling moxibustion.
MRATTS achieves recognition of all three moxibustion types by combining these metrics with critical threshold values derived from extensive testing.

\subsubsection{Visual Guidance}
\begin{figure}[htb]
  \centering
  \includegraphics[width=\linewidth]{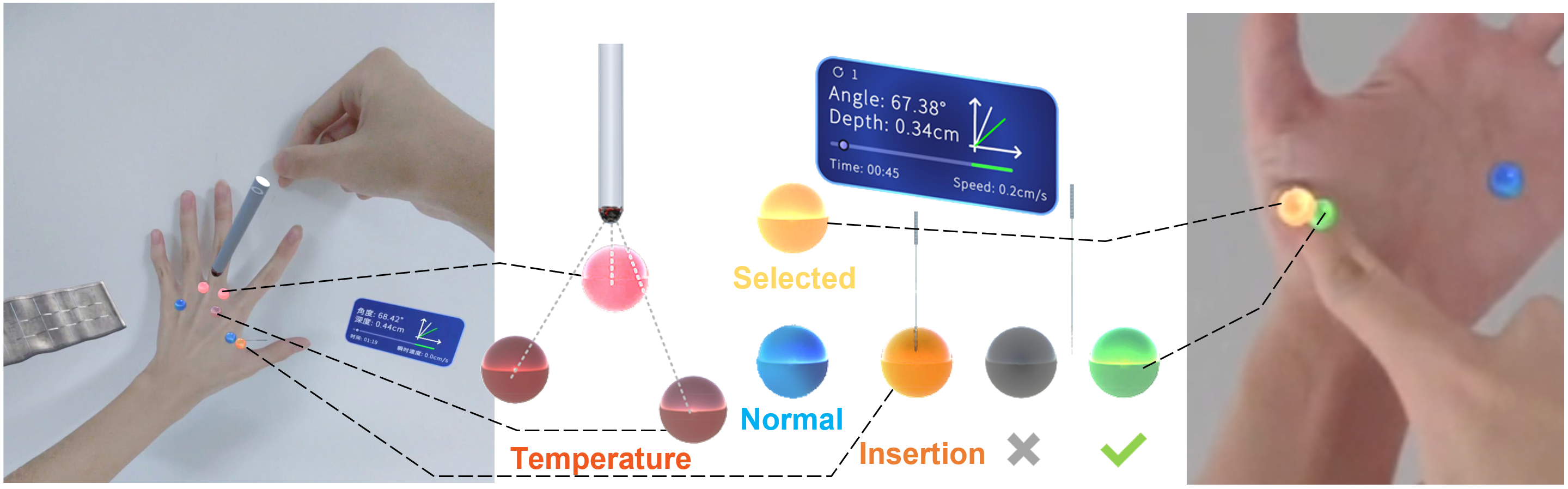}
  \caption{Visualization of acupoints in MRATTS. Acupoints are rendered as spheres with colors indicating the correctness of their locations (right), and insertion data is indicated in a separate panel (middle top). For moxibustion, colors are additionally used to encode temperature (left).}
  \label{fig:Acupoint color}
\end{figure}

We employ a semantic color-coding scheme to render acupoint spheres for acupressure, acupuncture, and moxibustion as shown in~\Cref{fig:Acupoint color}. 
Colors of spheres are dynamically updated to reflect the state of a procedure, supplemented by real-time data visualization to assist user training.

Acupressure: A small sphere is rendered at the pressing point of the index finger. Green is used to indicate a correct position, and gray for an incorrect one. This visual cue not only helps users gauge the spatial distance between the pressing point and the target acupoint but also indicates whether this distance falls within the valid effective range. Additionally, when an acupoint is selected for theoretical learning, it is highlighted in yellow.

Acupuncture: Upon successful needle insertion, the acupoint sphere transitions from blue to orange. Furthermore, key metrics, such as insertion angle and depth, are displayed in real-time during the procedure.

Moxibustion: A warm red spectrum is used to represent heat accumulation. The distance between the burning head of the moxa stick and all acupoints within a certain proximity is continuously calculated in real time. Based on this distance and a time-dependent function, the approximate skin temperature at the acupoint is determined, providing visual guidance through varying degrees/saturation of red coloring on the acupoint spheres.

\renewcommand{\arraystretch}{1.4}
\begin{table*}[p]
\centering
\caption{Scoring Categories and Functions}
\label{tab:variables}
\begin{tabular}{|c|cc|c|c|}
\hline
\textbf{Scoring Categories}                                            & \multicolumn{2}{c|}{\textbf{Scoring Function}}                                                                                                                                                             & \textbf{Symbol}                                   & \textbf{Explanation}                               \\ \hline
\rowcolor{amber!35} 
\cellcolor{amber!35}                                                   & \multicolumn{2}{c|}{\cellcolor{amber!35}}                                                                                                                                                                  & ${d_\mathrm{dev}}$                                & Deviation Distance                                 \\ \cline{4-5} 
\rowcolor{amber!35} 
\cellcolor{amber!35}                                                   & \multicolumn{2}{c|}{\cellcolor{amber!35}}                                                                                                                                                                  & $\mathrm{R_{min}}$                                & Full Score Range                                   \\ \cline{4-5} 
\rowcolor{amber!35} 
\multirow{-3}{*}{\cellcolor{amber!35}Acupressure}                      & \multicolumn{2}{c|}{\multirow{-3}{*}{\cellcolor{amber!35}Acupressure Location}}                                                                                                                            & $\mathrm{R_{max}}$                                & Zero Point Boundary                                \\ \hline
\rowcolor{orange!35} 
\cellcolor{orange!35}                                                  & \multicolumn{2}{c|}{\cellcolor{orange!35}}                                                                                                                                                                 & ${\theta_\mathrm{dev}}$                           & Needle Insertion Deviation Angle                   \\ \cline{4-5} 
\rowcolor{orange!35} 
\cellcolor{orange!35}                                                  & \multicolumn{2}{c|}{\cellcolor{orange!35}}                                                                                                                                                                 & $\mathrm{\uptheta_{min}}$                         & Full Score Range                                   \\ \cline{4-5} 
\rowcolor{orange!35} 
\cellcolor{orange!35}                                                  & \multicolumn{2}{c|}{\multirow{-3}{*}{\cellcolor{orange!35}Insertion Angle}}                                                                                                                                & $\mathrm{\uptheta_{max}}$                         & Zero Point Boundary                                \\ \cline{2-5} 
\rowcolor{orange!35} 
\cellcolor{orange!35}                                                  & \multicolumn{1}{c|}{\cellcolor{orange!35}}                                     & \cellcolor{orange!35}                                                                                                     & ${d_\mathrm{deep}}$                               & Deep Insertion Depth                               \\ \cline{4-5} 
\rowcolor{orange!35} 
\cellcolor{orange!35}                                                  & \multicolumn{1}{c|}{\cellcolor{orange!35}}                                     & \cellcolor{orange!35}                                                                                                     & $\mathrm{D^{lower}_{deep}}$                       & Deep Insertion Lower Bound                 \\ \cline{4-5} 
\rowcolor{orange!35} 
\cellcolor{orange!35}                                                  & \multicolumn{1}{c|}{\cellcolor{orange!35}}                                     & \cellcolor{orange!35}                                                                                                     & $\mathrm{D^{min}_{deep}}$                         & Deep Insertion Correct Range Minimum Value         \\ \cline{4-5} 
\rowcolor{orange!35} 
\cellcolor{orange!35}                                                  & \multicolumn{1}{c|}{\cellcolor{orange!35}}                                     & \cellcolor{orange!35}                                                                                                     & $\mathrm{D^{max}_{deep}}$                         & Deep Insertion Correct Range Maximum Value         \\ \cline{4-5} 
\rowcolor{orange!35} 
\cellcolor{orange!35}                                                  & \multicolumn{1}{c|}{\cellcolor{orange!35}}                                     & \multirow{-5}{*}{\cellcolor{orange!35}Deep Needling}        & $\mathrm{D^{upper}_{deep}}$                       & Deep Insertion Upper Bound                   \\ \cline{3-5} 
\rowcolor{orange!35} 
\cellcolor{orange!35}                                                  & \multicolumn{1}{c|}{\cellcolor{orange!35}}                                     & \cellcolor{orange!35}                                                                                                     & ${d_\mathrm{shallow}}$                            & Shallow Insertion Depth                            \\ \cline{4-5} 
\rowcolor{orange!35} 
\cellcolor{orange!35}                                                  & \multicolumn{1}{c|}{\cellcolor{orange!35}}                                     & \cellcolor{orange!35}                                                                                                     & ${D_{\text{shallow}}^{\text{lower}}}$             & Shallow Insertion Lower Bound              \\ \cline{4-5} 
\rowcolor{orange!35} 
\cellcolor{orange!35}                                                  & \multicolumn{1}{c|}{\cellcolor{orange!35}}                                     & \cellcolor{orange!35}                                                                                                     & $\mathrm{D^{min}_{shallow}}$                      & Shallow Insertion Correct Range Minimum Value      \\ \cline{4-5} 
\rowcolor{orange!35} 
\cellcolor{orange!35}                                                  & \multicolumn{1}{c|}{\cellcolor{orange!35}}                                     & \cellcolor{orange!35}                                                                                                     & $\mathrm{D^{max}_{shallow}}$                      & Shallow Insertion Correct Range Maximum Value      \\ \cline{4-5} 
\rowcolor{orange!35} 
\cellcolor{orange!35}                                                  & \multicolumn{1}{c|}{\cellcolor{orange!35}}                                     & \multirow{-5}{*}{\cellcolor{orange!35}Shallow Needling}     & $\mathrm{D^{upper}_{shallow}}$                    & Shallow Insertion Upper Bound                \\ \cline{3-5} 
\rowcolor{orange!35} 
\cellcolor{orange!35}                                                  & \multicolumn{1}{c|}{\cellcolor{orange!35}}                                     & \cellcolor{orange!35}                                                                                                     & $\mathrm{E^{shallow\_then\_deep}_{lift\_thrust}}$ & Shallow then Deep Event                            \\ \cline{4-5} 
\rowcolor{orange!35} 
\cellcolor{orange!35}                                                  & \multicolumn{1}{c|}{\cellcolor{orange!35}}                                     & \cellcolor{orange!35}                                                                                                     & $\mathrm{E^{deep\_then\_shallow}_{lift\_thrust}}$ & Deep then Shallow Event                            \\ \cline{4-5} 
\rowcolor{orange!35} 
\cellcolor{orange!35}                                                  & \multicolumn{1}{c|}{\cellcolor{orange!35}}                                     & \cellcolor{orange!35}                                                                                                     & $\mathrm{E^{fast\_then\_slow}_{lift\_thrust}}$    & Fast then Slow Event                               \\ \cline{4-5} 
\rowcolor{orange!35} 
\cellcolor{orange!35}                                                  & \multicolumn{1}{c|}{\cellcolor{orange!35}}                                     & \cellcolor{orange!35}                                                                                                     & $\mathrm{E^{slow\_then\_fast}_{lift\_thrust}}$    & Slow then Fast Event                               \\ \cline{4-5} 
\rowcolor{orange!35} 
\cellcolor{orange!35}                                                  & \multicolumn{1}{c|}{\cellcolor{orange!35}}                                     & \cellcolor{orange!35}                                                                                                     & $\mathrm{M^{reinforce}_{lift\_thrust}}$           & Lifting-Thrusting Reinforcing Method                \\ \cline{4-5} 
\rowcolor{orange!35} 
\cellcolor{orange!35}                                                  & \multicolumn{1}{c|}{\multirow{-16}{*}{\cellcolor{orange!35}Lifting-Thrusting}} & \cellcolor{orange!35}                                                                                                     & $\mathrm{M^{reduce}_{lift\_thrust}}$              & Lifting-Thrusting Reducing Method                   \\ \cline{4-5} 
\rowcolor{orange!35} 
\cellcolor{orange!35}                                                  & \multicolumn{1}{c|}{\cellcolor{orange!35}}                                     & \cellcolor{orange!35}                                                                                                     & $\mathrm{M^{target}_{lift\_thrust}}$              & Lifting-Thrusting Target Reinforcing/Reducing Method \\ \cline{4-5} 
\rowcolor{orange!35} 
\cellcolor{orange!35}                                                  & \multicolumn{1}{c|}{\cellcolor{orange!35}}                                     & \multirow{-8}{*}{\cellcolor{orange!35}Reinforcing-Reducing} & ${M^\mathrm{actual}_\mathrm{lift\_thrust}}$       & Lifting-Thrusting Actual Method \\ \cline{2-5} 
\rowcolor{orange!35} 
\cellcolor{orange!35}                                                  & \multicolumn{1}{c|}{\cellcolor{orange!35}}                                     & Twisting-turns Number                                                                                                     & ${n_\mathrm{total}}$                              & Accumulated Turns                                  \\ \cline{3-5} 
\rowcolor{orange!35} 
\cellcolor{orange!35}                                                  & \multicolumn{1}{c|}{\cellcolor{orange!35}}                                     & \cellcolor{orange!35}                                                                                                     & $\mathrm{E^{CW\_then\_CCW}_{twist}}$              & Clockwise then Counter-clockwise Event             \\ \cline{4-5} 
\rowcolor{orange!35} 
\cellcolor{orange!35}                                                  & \multicolumn{1}{c|}{\cellcolor{orange!35}}                                     & \cellcolor{orange!35}                                                                                                     & $\mathrm{E^{CCW\_then\_CW}_{twist}}$              & Counter-clockwise then Clockwise Event             \\ \cline{4-5} 
\rowcolor{orange!35} 
\cellcolor{orange!35}                                                  & \multicolumn{1}{c|}{\cellcolor{orange!35}}                                     & \cellcolor{orange!35}                                                                                                     & $\mathrm{M^{reinforce}_{twist}}$                  & Twisting Reinforcing Method                         \\ \cline{4-5} 
\rowcolor{orange!35} 
\cellcolor{orange!35}                                                  & \multicolumn{1}{c|}{\cellcolor{orange!35}}                                     & \cellcolor{orange!35}                                                                                                     & $\mathrm{M^{reduce}_{twist}}$                     & Twisting Reducing Method                            \\ \cline{4-5} 
\rowcolor{orange!35} 
\cellcolor{orange!35}                                                  & \multicolumn{1}{c|}{\cellcolor{orange!35}}                                     & \cellcolor{orange!35}                                                                                                     & $\mathrm{M^{target}_{twist}}$                     & Twisting Target Reinforcing/Reducing Method          \\ \cline{4-5} 
\rowcolor{orange!35} 
\multirow{-28}{*}{\cellcolor{orange!35}Acupuncture} & \multicolumn{1}{c|}{\multirow{-7}{*}{\cellcolor{orange!35}Twisting}}           & \multirow{-6}{*}{\cellcolor{orange!35}Reinforcing-Reducing} & ${M^\mathrm{actual}_\mathrm{twist}}$              & Twisting Actual Method          \\ \hline
\rowcolor{burntorange!35} 
\cellcolor{burntorange!35}                                             & \multicolumn{2}{c|}{\cellcolor{burntorange!35}}                                                                                                                                                            & ${d_\mathrm{moxi}}$                               & Moxibustion Distance                               \\ \cline{4-5} 
\rowcolor{burntorange!35} 
\cellcolor{burntorange!35}                                             & \multicolumn{2}{c|}{\cellcolor{burntorange!35}}                                                                                                                                                            & $\mathrm{D^{lower}_{moxi}}$                       & Moxibustion Lower Bound                     \\ \cline{4-5} 
\rowcolor{burntorange!35} 
\cellcolor{burntorange!35}                                             & \multicolumn{2}{c|}{\cellcolor{burntorange!35}}                                                                                                                                                            & $\mathrm{D^{min}_{moxi}}$                         & Moxibustion Correct Range Minimum Value            \\ \cline{4-5} 
\rowcolor{burntorange!35} 
\cellcolor{burntorange!35}                                             & \multicolumn{2}{c|}{\cellcolor{burntorange!35}}                                                                                                                                                            & $\mathrm{D^{max}_{moxi}}$                         & Moxibustion Correct Range Maximum Value            \\ \cline{4-5} 
\rowcolor{burntorange!35} 
\cellcolor{burntorange!35}                                             & \multicolumn{2}{c|}{\multirow{-5}{*}{\cellcolor{burntorange!35}Moxibustion Distance}}                                                                                                                      & $\mathrm{D^{upper}_{moxi}}$                       & Moxibustion Upper Bound                       \\ \cline{2-5} 
\rowcolor{burntorange!35} 
\cellcolor{burntorange!35}                                             & \multicolumn{2}{c|}{\cellcolor{burntorange!35}}                                                                                                                                                            & $\mathrm{T_{mild}}$                               & Mild Moxibustion                                   \\ \cline{4-5} 
\rowcolor{burntorange!35} 
\cellcolor{burntorange!35}                                             & \multicolumn{2}{c|}{\cellcolor{burntorange!35}}                                                                                                                                                            & $\mathrm{T_{sparrow}}$                            & Sparrow-pecking Moxibustion                        \\ \cline{4-5} 
\rowcolor{burntorange!35} 
\cellcolor{burntorange!35}                                             & \multicolumn{2}{c|}{\cellcolor{burntorange!35}}                                                                                                                                                            & $\mathrm{T_{whirling}}$                           & Whirling Moxibustion                               \\ \cline{4-5} 
\rowcolor{burntorange!35} 
\cellcolor{burntorange!35}                                             & \multicolumn{2}{c|}{\cellcolor{burntorange!35}}                                                                                                                                                            & $\mathrm{T_{target}}$                             & Target Type                                        \\ \cline{4-5} 
\rowcolor{burntorange!35} 
\multirow{-10}{*}{\cellcolor{burntorange!35}Moxibustion} & \multicolumn{2}{c|}{\multirow{-5}{*}{\cellcolor{burntorange!35}Moxibustion Type}}                                                                                                                          & ${T_\mathrm{actual}}$                             & Actual Type                                        \\ \hline
\end{tabular}
\end{table*}

\subsection{Evaluation Standards and User Performance Report}
\label{sec:Evaluation standards}
Post-practice evaluation and feedback are a critical phase in the acupoint therapy training workflow. As detailed in ~\Cref{sec:Acupoint knowledge},~\Cref{sec:Simulated practice and visual guidance}, and \autoref{tab:knowledge}, TCM acupoint therapy involves complex procedures, each consisting of multiple distinct steps. Therefore, a set of comprehensive step-by-step evaluation standards with grading extending the binary correct/incorrect judgment is required. 
The grading should be interpretable to provide users with fine-grained feedback on their performance, specifically highlighting the degree of error.

We collaborate with E1 to develop such a set of evaluation standards.  
These standards are grounded in descriptions of valid therapeutic ranges found in TCM classics ~\cite{GBT21709.21,GBT21709.20,TCAAM001,GBT21709.1} and extensive clinical experience of practitioners.   
They provide qualitative and quantitative scoring based on the therapeutic efficacy and risk coefficients associated with the user's operational data recorded during the simulated practice session.
\begin{figure*}[tb]
  \centering 
  \includegraphics[width=1\linewidth]{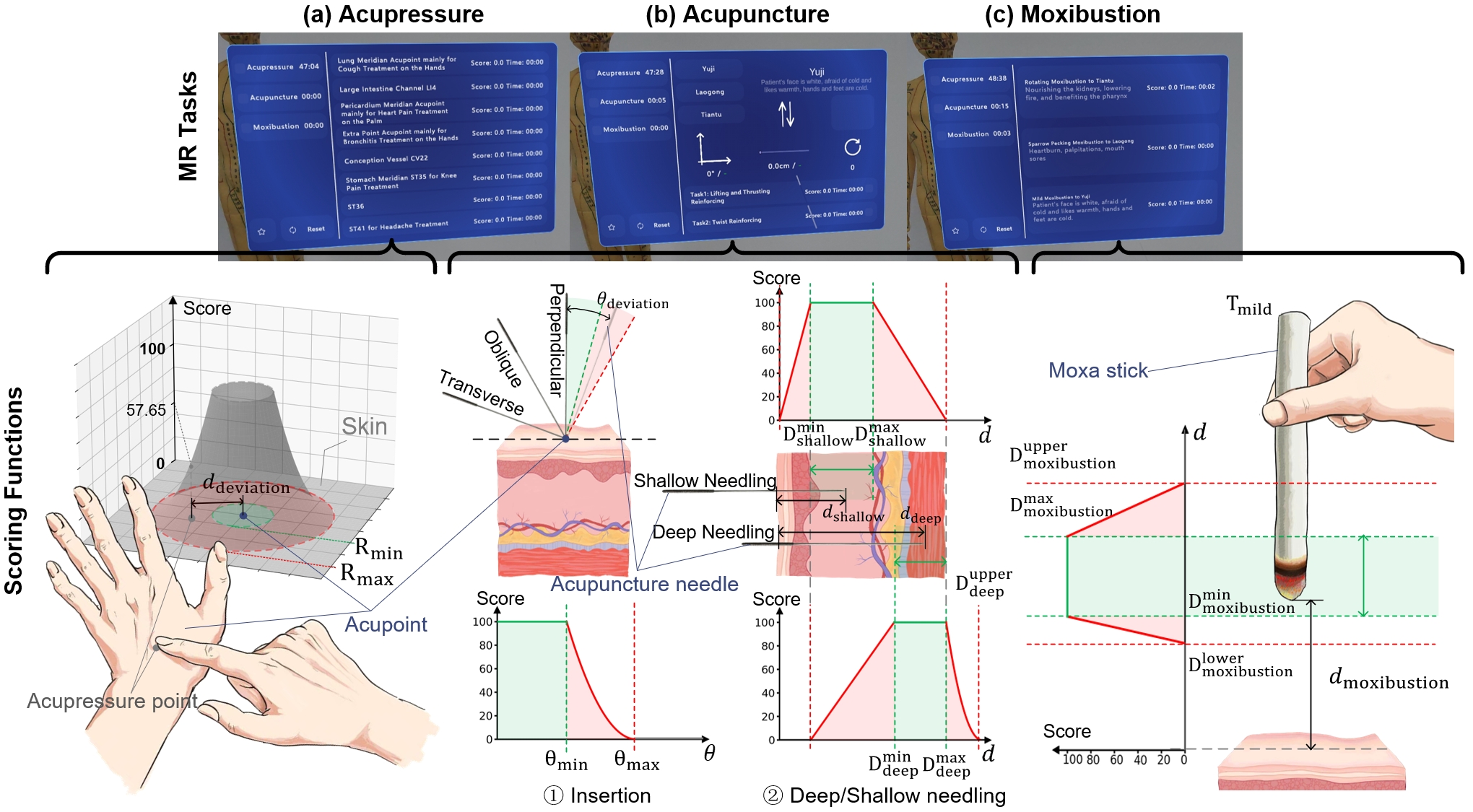}
  \caption{The evaluation standards for (a) acupressure, (b) acupuncture, and (c) moxibustion. The task panels (top) and the scoring functions (bottom) are shown for each technique, respectively. 
  }
  \vspace{-1em}
  \label{fig:evalFuncsTask} 
\end{figure*}
We integrate the scores with the visualizations of operational data to generate a user performance report, as shown in \Cref{fig:evalFuncsTask}(top). The variables below are described in \autoref{tab:variables}.

\subsubsection{Acupressure}
Acupressure is simple and widely used, and is characterized by relatively straightforward manipulation techniques. Therefore, our evaluation standards focus primarily on the learner's acupoint localization ability. 
This ability is the cornerstone of acupoint therapy, as all treatments rely on correct acupoint location. The acupoint is not a single point but rather a certain area~\cite{li2013big}. Some studies have shown that the effective area of acupoints can be large~\cite{li2015acupoint,kong2009functional}. We aim to help users learn the most accurate acupoint location according to the theory~\cite{world2008standard,GBT12346-2021,wu2022interpretation,GBT40997-2021}.
 
 We specify a correct circular region with a radius $\mathrm{R}_\mathrm{min}$ and an invalid boundary of $\mathrm{R}_\mathrm{max}$ (indicating negligible therapeutic effect). The specific values of these radii depend on the anatomical location of the acupoint. For instance, the effective area of acupoints on the hands is significantly smaller than that on the torso, necessitating higher localization precision. We define the offset distance $\mathit{d}_{\mathrm{dev}}$ between the location determined by the user's pressing operation and the location automatically detected by our method. When $\mathit{d}_\mathrm{dev}$ is within $\mathrm{R}_\mathrm{min}$, the full points are achieved, while outside $\mathrm{R}_\mathrm{max}$, the score is zero.
 
 A quadratic decrease in the $\mathit{d}_{\mathrm{dev}}$ score within the range $[\mathrm{R}_\mathrm{min}, \mathrm{R}_\mathrm{max}]$ is applied to penalize users for deviating more from the correct region
 as shown in \Cref{fig:evalFuncsTask}(a) and \Cref{equ:AL}.

\begin{equation}
    \renewcommand{\arraystretch}{1.5}
    \scalebox{0.87}{$
    \mathrm{AL}(\bm{d}_{\mathbf{dev}}) = 
    \left\{
    \begin{array}{cl}
    \displaystyle \mathrm{C}  & \bm{d}_{\mathbf{dev}} < \mathrm{R}_{\mathrm{min}} \\[10pt]
    \displaystyle \frac{\mathrm{C} \cdot (\bm{d}_{\mathbf{dev}} - \mathrm{R}_{\mathrm{max}})^2}{(\mathrm{R}_{\mathrm{min}} - \mathrm{R}_{\mathrm{max}})^2} &  \mathrm{R}_{\mathrm{min}} \leq  \bm{d}_{\mathbf{dev}} < \mathrm{R}_{\mathrm{max}} \\[10pt]
    0 & \bm{d}_{\mathbf{dev}} \geq \mathrm{R}_{\mathrm{max}}
    \end{array} 
    \right.
    $}
    \label{equ:AL}
\end{equation}

\subsubsection{Acupuncture}
The evaluation of acupuncture manipulations focuses on three techniques: insertion, lifting-thrusting, and twisting. Furthermore, the assessment of therapeutic efficacy incorporates the correct application of reinforcing-reducing methods.
Specific scoring strategies are designed for the distinct characteristics of these techniques.

\textbf{Insertion:} the range of insertion angle is used for classification. A quantitative scoring function on angles is designed (\Cref{fig:evalFuncsTask}(b)).
Expert E1 emphasizes that insertion is the foundation of acupuncture, and correct insertion angles are a prerequisite for subsequent lifting-thrusting and twisting manipulations.

A piecewise function is used to determine the score.
A deviation ${\theta_\mathrm{dev}}$ within $\uptheta_\mathrm{min}$ of the target value is considered a minor error and earns full points. When the deviation exceeds $\uptheta_\mathrm{min}$, we apply a quadratic decrease to increase the penalty and provide stronger negative feedback similar to that in the acupressure module.
Conversely, any deviation beyond $\uptheta_\mathrm{max}$ is considered an operational error resulting in a score of zero, as shown in \Cref{equ:AN} and \Cref{fig:evalFuncsTask}(b)~\circled{1}. 
$\uptheta_\mathrm{max}$ is recommended to be the midpoint between the angles of adjacent techniques to serve as a threshold for distinguishing potential confusion.  
Meanwhile, $\uptheta_\mathrm{min}$ can be customized by the user according to their training objectives (i.e., the required precision of insertion).

\begin{equation}
    \renewcommand{\arraystretch}{1.5} 
    \scalebox{0.87}{$
    \mathrm{IA}(\bm{\theta}_{\mathbf{dev}}) = 
    \left\{
    \begin{array}{cl}
    \mathrm{C} & 0 \leq \bm{\theta}_{\mathbf{dev}} < \mathrm{\uptheta}_{\mathrm{min}} \\[12pt]
    \displaystyle \frac{\mathrm{C} \cdot (\bm{\theta}_{\mathbf{dev}} - \mathrm{\uptheta}_{\mathrm{max}})^2}{(\mathrm{\uptheta}_{\mathrm{min}} - \mathrm{\uptheta}_{\mathrm{max}})^2} &  \mathrm{\uptheta}_{\mathrm{min}}  \leq \bm{\theta}_{\mathbf{dev}} < \mathrm{\uptheta}_{\mathrm{max}} \\[12pt]
    0 & \bm{\theta}_{\mathbf{dev}} \geq \mathrm{\uptheta}_{\mathrm{max}}
    \end{array}
    \right.
    $}
    \label{equ:AN}
\end{equation}

\begin{figure*}[tb]
  \centering 
  \includegraphics[width=0.95\linewidth]{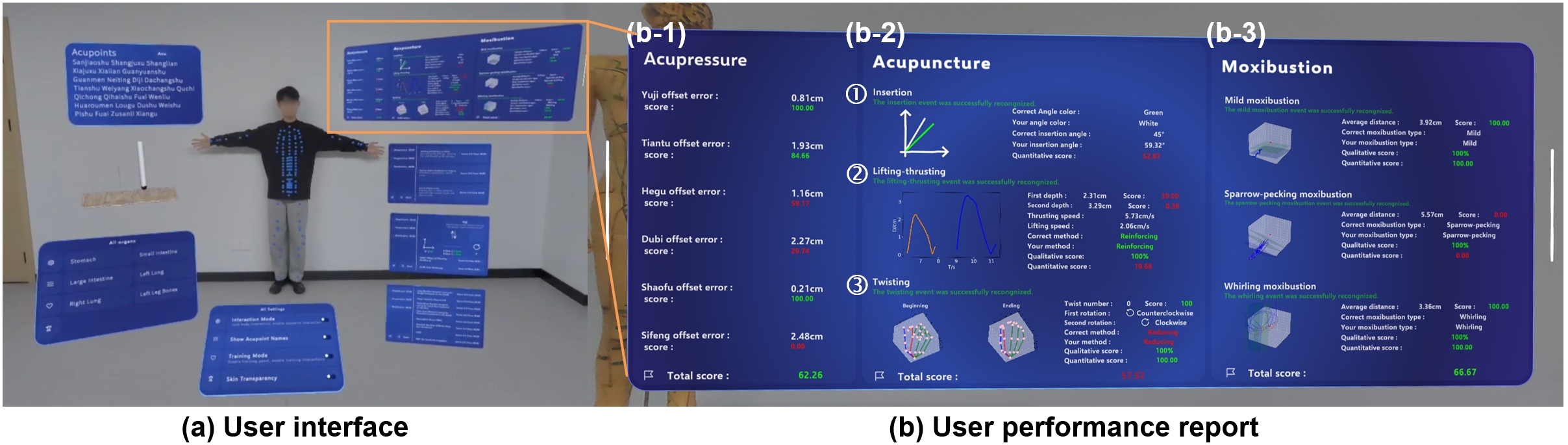}\vspace{-0.5em}
  \caption{The (a) full user interface of MRATTS comprises multiple panels, including (b) the performance report.
  }
  \label{fig:MR Evaluation}
\end{figure*}

The visualization of insertion performance is shown in \Cref{fig:MR Evaluation}(b-2)~\circled{1}: the green line represents the correct insertion angle, while the white line indicates the simulated insertion angle of the user, providing intuitive feedback on the deviation between them.

\textbf{Lifting-thrusting:} the core criteria involve both the depth range and the specific reinforcing-reducing type. A hybrid quantitative and qualitative scoring approach is adopted (\Cref{fig:evalFuncsTask}(b)~\circled{2}).

The quantitative scoring focuses on the judgment of depth for both deep and shallow needling, as shown in \Cref{fig:evalFuncsTask}(b)~\circled{2}. 
When the deep insertion ${d}_{\mathrm{deep}}$ is too shallow (${d}_{\mathrm{deep}}<{\mathrm{D}}_{\mathrm{deep}}^{\mathrm{min}}$) or the shallow insertion ${d}_{\mathrm{shallow}}$ is too shallow (${d}_{\mathrm{shallow}}<{\mathrm{D}}_{\mathrm{shallow}}^{\mathrm{min}}$), it generally does not pose a significant safety risk but affects the therapeutic efficacy. 
Therefore, after deviating from the correct range ($[{\mathrm{D}}_{\mathrm{shallow}}^{\mathrm{min}},{\mathrm{D}}_{\mathrm{shallow}}^{\mathrm{max}}]$ and $[{\mathrm{D}}_{\mathrm{deep}}^{\mathrm{min}},{\mathrm{D}}_{\mathrm{deep}}^{\mathrm{max}}]$), the score decreases linearly until the operation is deemed erroneous and reaches 0. 
However, when the deep insertion is excessively deep (${d}_{\mathrm{deep}}\geq{\mathrm{D}}_{\mathrm{deep}}^{\mathrm{max}}$), it poses a threat to patient safety~\cite{chou2011safe}, the score decreases quadratically until it reaches zero at ${\mathrm{D}}_{\mathrm{deep}}^{\mathrm{upper}}$ (\Cref{equ:DN}). 
To avoid mistakenly identifying a shallow insertion that goes too deep as a deep insertion, 
${\mathrm{D}}_{\mathrm{shallow}}^{\mathrm{upper}}$ is set to be equal to ${\mathrm{D}}_{\mathrm{deep}}^{\mathrm{max}}$. 
Similarly, the zero-score boundary ${\mathrm{D}}_{\mathrm{deep}}^{\mathrm{lower}}$ for deep insertion is set to ${\mathrm{D}}_{\mathrm{shallow}}^{\mathrm{min}}$ as shown in \Cref{fig:evalFuncsTask}(b)~\circled{2}.

\begin{equation}
    \renewcommand{\arraystretch}{1.5} 
    \scalebox{0.87}{$
    \mathrm{DN}(\bm{d}_{\mathbf{deep}}) = 
    \left\{
    \begin{array}{cl}
    0 &  0 \leq \bm{d}_{\mathbf{deep}} < \mathrm{D}_{\mathrm{deep}}^{\mathrm{lower}} \\[12pt]
    \displaystyle \frac{\mathrm{C} \cdot (\bm{d}_{\mathbf{deep}} - \mathrm{D}_{\mathrm{deep}}^{\mathrm{lower}})}{\mathrm{D}^{\mathrm{min}}_{\mathrm{deep}} - \mathrm{D}_{\mathrm{deep}}^{\mathrm{lower}}} & \mathrm{D}_{\mathrm{deep}}^{\mathrm{lower}} \leq \bm{d}_{\mathbf{deep}} < \mathrm{D}^{\mathrm{min}}_{\mathrm{deep}} \\[12pt]
    \mathrm{C} &  \mathrm{D}^{\mathrm{min}}_{\mathrm{deep}} \leq \bm{d}_{\mathbf{deep}} < \mathrm{D}^{\mathrm{max}}_{\mathrm{deep}} \\[12pt]
    \displaystyle \frac{\mathrm{C} \cdot (\bm{d}_{\mathbf{deep}} - \mathrm{D}_{\mathrm{deep}}^{\mathrm{upper}})^2}{(\mathrm{D}_{\mathrm{deep}}^{\mathrm{max}} - \mathrm{D}_{\mathrm{deep}}^{\mathrm{upper}})^2} &  \mathrm{D}^{\mathrm{max}}_{\mathrm{deep}} \leq \bm{d}_{\mathbf{deep}} < \mathrm{D}_{\mathrm{deep}}^{\mathrm{upper}} \\[12pt]
    0 & \bm{d}_{\mathbf{deep}} \geq \mathrm{D}_{\mathrm{deep}}^{\mathrm{upper}} 
    \end{array}
    \right.
    $}
    \label{equ:DN}
\end{equation}

\begin{equation}
    \renewcommand{\arraystretch}{1.5}
    \scalebox{0.78}{$
    \mathrm{SN}(\bm{d}_{\mathbf{shallow}}) = 
    \left\{
    \begin{array}{cl}
    \displaystyle \frac{\mathrm{C} \cdot \bm{d}_{\mathbf{shallow}}}{\mathrm{D}_{\mathrm{shallow}}^{\mathrm{min}}}
     & \mathrm{D}_{\mathrm{shallow}}^{\mathrm{lower}} \leq \bm{d}_{\mathbf{shallow}} < \mathrm{D}_{\mathrm{shallow}}^{\mathrm{min}} \\[12pt]
    \mathrm{C} & \mathrm{D}_{\mathrm{shallow}}^{\mathrm{min}} \leq \bm{d}_{\mathbf{shallow}} < \mathrm{D}_{\mathrm{shallow}}^{\mathrm{max}} \\[12pt]
    \displaystyle \frac{\mathrm{C} \cdot \left( \bm{d}_{\mathbf{shallow}} - \mathrm{D}_{\mathrm{shallow}}^{\mathrm{over}} \right)}{\mathrm{D}_{\mathrm{shallow}}^{\mathrm{max}} - \mathrm{D}_{\mathrm{shallow}}^{\mathrm{over}}}
     & \mathrm{D}_{\mathrm{shallow}}^{\mathrm{max}} \leq \bm{d}_{\mathbf{shallow}} < \mathrm{D}_{\mathrm{shallow}}^{\mathrm{upper}} \\[12pt]
    0 & \bm{d}_{\mathbf{shallow}} \geq \mathrm{D}_{\mathrm{shallow}}^{\mathrm{upper}} 
    \end{array}
    \right.
    $}
    \label{equ:SN}
\end{equation}

Even if the quantitative deep and shallow needling performance is correct, the therapeutic efficacy is regarded as reduced if the operational sequence is reversed. 
Therefore, we design the qualitative scoring of reinforcing-reducing (\Cref{sec:acupuncture}) as a weighting factor. 
When the actual sequence $\mathit{M}_{\mathrm{lift\_thrust}}^{\mathrm{actual}}$ of the user matches the target method $\mathrm{M}_{\mathrm{lift\_thrust}}^{\mathrm{target}}$, a full weight of 100\% is awarded. Otherwise, the weight is reduced to 60\% (\Cref{equ:LT_RR}).

\begin{equation}
    \renewcommand{\arraystretch}{1.5}
    \scalebox{0.95}{$ 
    \mathrm{M}_{\mathrm{lift\_thrust}}^{\mathrm{reinforce}} = 
    \mathrm{E}_{\mathrm{lift\_thrust}}^{\mathrm{deep\_then\_shallow}} \cap \mathrm{E}_{\mathrm{lift\_thrust}}^{\mathrm{fast\_then\_slow}}
    $}
    \label{equ:LT-rein}
\end{equation}

\begin{equation}
    \renewcommand{\arraystretch}{1.5}
    \scalebox{0.95}{$ 
    \mathrm{M}_{\mathrm{lift\_thrust}}^{\mathrm{reduce}} = 
    \mathrm{E}_{\mathrm{lift\_thrust}}^{\mathrm{shallow\_then\_deep}} \cap \mathrm{E}_{\mathrm{lift\_thrust}}^{\mathrm{slow\_then\_fast}}
    $}
    \label{equ:LT-redu}
\end{equation}

\begin{equation}
    \renewcommand{\arraystretch}{1.5}
    \scalebox{0.95}{$ 
    \mathrm{M}_{\mathrm{lift\_thrust}}^{\mathrm{target}} \in\left\{\mathrm{M}_{\mathrm{lift\_thrust}}^{\mathrm{reinforce}},
    \mathrm{M}_{\mathrm{lift\_thrust}}^{\mathrm{reduce}}\right\}
    $}
\end{equation}

\begin{equation}
    \renewcommand{\arraystretch}{1.5}
    \scalebox{0.85}{$ 
    \mathrm{LT\_RR}(\bm{M}_{\mathbf{lift\_thrust}}^{\mathbf{actual}}) = 
    \begin{cases} 
    100\% &  \bm{M}_{\mathbf{lift\_thrust}}^{\mathbf{actual}} = \mathrm{M}_{\mathrm{lift\_thrust}}^{\mathrm{target}} \\
    60\% &  \bm{M}_{\mathbf{lift\_thrust}}^{\mathbf{actual}} \neq \mathrm{M}_{\mathrm{lift\_thrust}}^{\mathrm{target}}
    \end{cases}
    $}
    \label{equ:LT_RR}
\end{equation}

The total score for lifting-thrusting $S_{\mathrm{L-T}}$ is then the quantitative score  (\Cref{equ:DN,equ:SN}) scaled by the qualitative coefficient $\mathrm{LT\_RR}$ (\Cref{equ:LT_RR}):
\begin{equation}
    \scalebox{0.98}{$
    S_{\mathrm{L-T}} =
  \frac{\mathrm{DN}(d_{\mathrm{deep}}) + \mathrm{SN}(d_{\mathrm{shallow}})}{2}\cdot  \mathrm{LT\_RR}(M_{\mathrm{lift\_thrust}}^{\mathrm{actual}})  
    $}\;.
\label{eq:lifting_thrusting_total}
\end{equation}

The visualization of lifting-thrusting performance is shown in \Cref{fig:MR Evaluation}(b-2)~\circled{2}. 
Our objective is to allow users to clearly observe the sequential relationship between their deep and shallow insertions, as well as the variations in operation speed. 
Therefore, a depth-vs-time plot is used: by observing depth on the y-axis against the temporal progression on the x-axis, users can verify their operation sequence. 
Furthermore, the slope of the curve intuitively depicts the speed contrast between lifting and thrusting.
    
\textbf{Twisting:} the number of clockwise and counter-clockwise rotations should remain balanced by the end of the operation. Quantitative scoring is based on the rotation count. Meanwhile, the direction of twisting determines the distinction between reinforcing-reducing methods; qualitative scoring is employed for this aspect.

The quantitative scoring $\mathrm{TN}$ for twisting checks if the needle fails to return to its original orientation upon completion (i.e., the net rotation derived from clockwise and counter-clockwise movements is not zero) as it may cause tissue damage: such instances result in a score of zero, the full score is otherwise awarded (\Cref{equ:TN}).

\begin{equation}
    \renewcommand{\arraystretch}{1.5} 
    \scalebox{0.95}{$ 
        \mathrm{TN}(\bm{n}_{\mathbf{total}}) = 
        \left\{
        \begin{array}{cl}
            100\% &  \bm{n}_{\mathbf{total}} = 0 \\
            0\% &  \bm{n}_{\mathbf{total}} \neq 0
        \end{array} 
        \right.
    $}
    \label{equ:TN}
\end{equation}

The qualitative scoring $\mathrm{T\_RR}$ is the same as that of the lifting-thrusting: if the actual operation matches the target method, a full weight of 100\% is awarded; otherwise, the weight is reduced to 60\% (\Cref{equ:T_RR}).

\begin{equation}
\renewcommand{\arraystretch}{1.5}
\scalebox{0.95}{$
\mathrm{M}_{\mathrm{twist}}^{\mathrm{reinforce}} = \mathrm{E}_{\mathrm{twist}}^{\mathrm{CW\_then\_CCW}}, \quad 
\mathrm{M}_{\mathrm{twist}}^{\mathrm{reduce}} = \mathrm{E}_{\mathrm{twist}}^{\mathrm{CCW\_then\_CW}}
$}
\label{equ:T-re}
\end{equation}

\begin{equation}
\renewcommand{\arraystretch}{1.5}
\scalebox{0.95}{$
\mathrm{M}_{\mathrm{twist}}^{\mathrm{target}} 
\in \left\{\mathrm{M}_{\mathrm{twist}}^{\mathrm{reinforce}},
\mathrm{M}_{\mathrm{twist}}^{\mathrm{reduce}} \right\}
$}
\label{equ:M_target}
\end{equation}

\begin{equation}
\renewcommand{\arraystretch}{1.5}
\scalebox{0.95}{$
\mathrm{T\_RR}(\bm{M}_{\mathbf{twist}}^{\mathbf{actual}}) = 
\begin{cases} 
100\% &  \bm{M}_{\mathbf{twist}}^{\mathbf{actual}} = \mathrm{M}_{\mathrm{twist}}^{\mathrm{target}} \\
60\% &  \bm{M}_{\mathbf{twist}}^{\mathbf{actual}} \neq \mathrm{M}_{\mathrm{twist}}^{\mathrm{target}}
\end{cases}
$}
\label{equ:T_RR}
\end{equation}

Therefore, the total score for twisting $S_{\mathrm{Twist}}$ is the quantitative score (\Cref{equ:TN}) scaled by the qualitative weighting factor (\Cref{equ:T_RR}):
\begin{equation}
    S_{\mathrm{Twist}} = 
    \mathrm{TN}(n_{\mathrm{total}}) \cdot \mathrm{T\_RR}(M_{\mathrm{twist}}^{\mathrm{actual}})\;.
    \label{eq:twisting_total}
\end{equation}

The visualization of twisting data is designed for clear identification of the rotation direction.  
Hand joint positions at the beginning and the end of the first twisting maneuver are visualized as shown in \Cref{fig:MR Evaluation}(b-2)~\circled{3}.  This allows users to observe whether the first rotation is clockwise or counter-clockwise, thereby facilitating the determination of the reinforcing or reducing method.

\subsubsection{Moxibustion}
The distance between the tip of the moxa stick and the skin is crucial as moxibustion works primarily through thermal stimulation---generated by the combustion of mugwort within the moxa stick---on acupoints~\cite{TCAAM0013}.
However, if the manipulation is inadequate, e.g., a low degree of completion or incorrect moxibustion type, the intended therapeutic effects may not be achieved.
Therefore, we also require the scoring to represent the completion status of the specific moxibustion type.

Evidence suggests that the optimal distance for moxibustion is approximately $3\,\mathrm{cm}$ from the skin, and is affected by the environmental temperature~\cite{liu2022numerical,nakamura2011noninvasive}. 
We set extreme values for the optimal moxibustion distance under different environmental temperatures as the correct range for moxibustion. 
During moxibustion, if the distance is less than $2\,\mathrm{cm}$, the patient may feel a burning sensation and may get injured. 
Therefore, we set the value of $\mathrm{D_{moxi}^{min}} = 2.5\,\mathrm{cm}$ and $\mathrm{D_{moxi}^{max}} = 4\,\mathrm{cm}$, and moxibustion operations within the range [$\mathrm{D_{moxi}^{min}}$, $\mathrm{D_{moxi}^{max}}$] get full points. 
Deviating from this range results in linearly decreasing scores until reaching $\mathrm{D_{moxi}^{upper}} = 5\,\mathrm{cm}$ and $\mathrm{D_{moxi}^{lower}} = 2\,\mathrm{cm}$, where no point is awarded. 
The distance-based scoring function $\mathrm{MD}$ is shown in \Cref{fig:evalFuncsTask}(c) and \Cref{equ:MD}.

\begin{equation}
\renewcommand{\arraystretch}{1.5}
\scalebox{0.79}{$
\mathrm{MD}(\bm{d_{\mathbf{moxi}}}) = 
\left\{
\begin{array}{cl}
0 & 0 \leq \bm{d_{\mathbf{moxi}}} < \mathrm{D_{\mathit{moxi}}^{lower}} \\[10pt]
\displaystyle \frac{\mathrm{C}\cdot ( \bm{d_{\mathbf{moxi}}} - \mathrm{D_{\mathrm{moxi}}^{lower}} )^2}{(\mathrm{D_{\mathrm{moxi}}^{min}} - \mathrm{D_{\mathrm{moxi}}^{lower}})^2} & \mathrm{D_{\mathrm{moxi}}^{lower}} \leq \bm{d_{\mathbf{moxi}}} < \mathrm{D_{\mathrm{moxi}}^{min}} \\[10pt]
\mathrm{C} & \mathrm{D_{\mathrm{moxi}}^{min}} \leq \bm{d_{\mathbf{moxi}}} < \mathrm{D_{\mathrm{moxi}}^{max}} \\[10pt]
\displaystyle \frac{\mathrm{C}\cdot ( \bm{d_{\mathbf{moxi}}} - \mathrm{D_{\mathrm{moxi}}^{upper}} )}{\mathrm{D_{\mathrm{moxi}}^{max}} - \mathrm{D_{\mathrm{moxi}}^{upper}}} & \mathrm{D_{\mathrm{moxi}}^{max}} \leq \bm{d_{\mathbf{moxi}}} < \mathrm{D_{\mathrm{moxi}}^{upper}} \\[10pt]
0 & \bm{d_{\mathbf{moxi}}} \geq \mathrm{D_{\mathrm{moxi}}^{upper}}
\end{array} 
\right.
$}
\label{equ:MD}
\end{equation}

A weighting coefficient $\mathrm{MT}$ similar to that of acupuncture is used: if the moxibustion type $T_{\mathrm{actual}}$ performed by the user matches the target type $\mathrm{T}_{\mathrm{target}}$, a full weight of 100\% is awarded. Otherwise, the weight is reduced to 60\%.

\begin{equation}
\renewcommand{\arraystretch}{1.5}
\scalebox{0.95}{$
    \mathrm{T}_{\mathrm{target}} \in \left\{ \mathrm{T}_{\mathrm{sparrow}}, \mathrm{T}_{\mathrm{whirling}}, \mathrm{T}_{\mathrm{mild}} \right\}
$}
\label{equ:T_target}
\end{equation}

\begin{equation}
\renewcommand{\arraystretch}{1.5}
\scalebox{0.95}{$
    \mathrm{MT}(\bm{T}_{\mathbf{actual}}) = 
\left\{
\begin{array}{cl}
100\% &  \bm{T}_{\mathbf{actual}} = \mathrm{T}_{\mathrm{target}} \\
60\% &  \bm{T}_{\mathbf{actual}} \neq \mathrm{T}_{\mathrm{target}}
\end{array} 
\right.
$}
\label{eq:MT}
\end{equation}

The total moxibustion score $S_{\mathrm{Moxi}}$ is then calculated as: 
\begin{equation}
    S_{\mathrm{Moxi}} = \mathrm{MD}(d_{\mathrm{moxi}}) \cdot \mathrm{MT}(T_{\mathrm{actual}})\;.
    \label{eq:moxibustion_total}
\end{equation}

After the moxibustion process, the trajectory of the moxa stick is visualized as shown in \Cref{fig:MR Evaluation}(b-3). The color intensity along the trajectory represents the instantaneous movement speed. Orange markers highlight key points identified by conditional logic, which serve to differentiate between sparrow-pecking and whirling moxibustion.

\subsection{User Interface and Implementation}    
The full user interface of MRATTS is shown in \Cref{fig:MR Evaluation}(a). The recipient of the simulated practice is overlaid by detected acupoints (\Cref{sec:Acupoint detection and visualization}). The virtual objects of acupuncture needles and moxi sticks are rendered at the center left.
Several panels are positioned in the virtual environment, including the acupoint knowledge panel (top left), the acupoint therapy tasks (bottom right), and the performance report (top right,~\Cref{sec:Evaluation standards}).

MRATTS was implemented using Unity3D~\cite{xie2012research} with the Mixed Reality Toolkit (MRTK)~\cite{mrtk2015,picomrtk3}, while the image was obtained through the RGB camera of the PICO 4 Pro.
To evaluate the effectiveness and usefulness of MRATTS, we conducted numerical evaluations (\Cref{sec:numericalStudy}) and user-oriented studies (\Cref{sec:userStudies}).

\section{Numerical Evaluation}
\label{sec:numericalStudy}
The performance of MRATTS is evaluated through numerical experiments of acupoint detection accuracy and runtime performance analysis.
Experiments were performed in our laboratory using a PICO 4 Pro HMD and a PC workstation with an Intel i7-11800H CPU, 16 GB of RAM, and NVIDIA RTX 3060 GPU with 6 GB memory. 

\subsection{Detection Accuracy Assessment}

We assess the system's detection accuracy by comparing the predicted acupoint locations against expert-annotated ground truth. 

\begin{figure}[h]
      \centering 
      \includegraphics[width=0.9\linewidth]{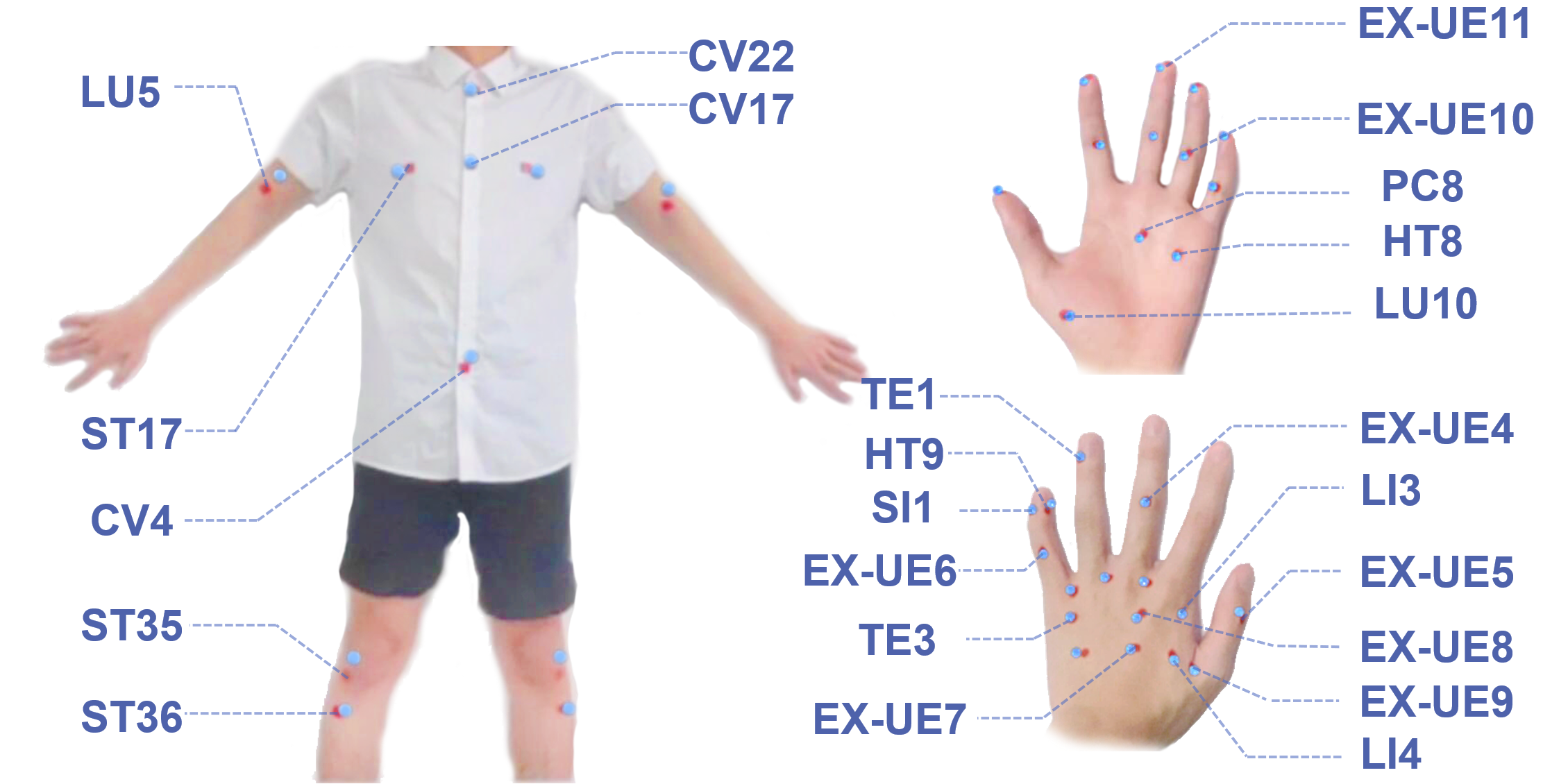}
      \caption{
     Visualization of the representative acupoints we selected.
      }
      \label{fig:acupoint sys and prof}
    \end{figure}

It is important to note that not all acupoints are suitable for visual detection. While most body acupoints can be located using the B-cun and F-cun methods, others rely on palpable surface anatomical landmarks. Therefore, based on professional TCM theories~\cite{wu2022interpretation,GBT40997-2021,lim2010standard,world2008standard,GBT12346-2021}, we first selected a subset of representative acupoints compatible with B-cun and F-cun methods. The representative acupoints include (\Cref{fig:acupoint sys and prof}):

    \begin{itemize}
        \item Hand: LU10, HT8, PC8, EX-UE10, EX-UE11, SI1, HT9, TE1, TE3, EX-UE5, LI3, LI4, EX-UE9, EX-UE7, EX-UE4, EX-UE8, EX-UE6.
        \item Torso and Limbs: CV22, CV17, ST17, ST35, ST36, ST41, LU5. 
    \end{itemize}

Then, we marked their physical locations on human subjects under the guidance of E1. These markings were reviewed and verified by E2---an experienced acupoint therapy expert who was an associate chief physician with 12 years of clinical experience. Specifically, $1\,\mathrm{cm} \times 1\,\mathrm{cm}$ red stickers were used for the limbs and torso, while a red marker was used to dot the locations on the hands. The automatically detected acupoints are also overlaid for comparison. Then, we evaluated the errors between the marked locations $\boldsymbol{P}_\text{marked}$ and the detected results $\boldsymbol{P}_\text{pred}$ for individuals with different body types (including overweight and tall subjects).

Finally, we analyze the deviations between $\boldsymbol{P}_\text{marked}$ and $\boldsymbol{P}_\text{pred}$. We manually reviewed and selected video segments from the experimental recordings where human recipients exhibited either slow movements or complete stillness. 
Subsequently, 95 frames are randomly selected from these segments to calculate the relative error and absolute error.

To account for the variation in the proportion of pixels occupied by a person across different frames, raw pixel Euclidean distance ($\Delta{p}$ between points $\boldsymbol{P}_\text{pred}$ and $\boldsymbol{P}_\text{marked}$, as shown in \Cref{equ:p}) lacks universal significance. Therefore, we introduce a normalization constant $D$ to compute the Relative Error (RE). Furthermore, to evaluate the physical deviation, we convert this to Absolute Error (AE), measured in centimeters. The calculation logic is defined in \Cref{equ:re}:

\begin{eqnarray}
    &\Delta{p} = \|\boldsymbol{P}_\text{pred} - \boldsymbol{P}_\text{marked}\|
    \label{equ:p}\;, \\
    &\mathrm{RE} = \frac{\Delta{p}}{{D}},   \mathrm{AE} = \mathrm{RE} \cdot {L}
    \label{equ:re} \;.
\end{eqnarray}

    We select different normalization constants $D$ for different body parts to ensure scale invariance:

    \begin{itemize}

        \item For Hand Acupoints: $D$ is defined as the pixel distance from EX-UE10 to EX-UE11 (the distal phalanx of the middle finger).
    
        \item For Limb and Torso Acupoints: $D$ is defined as the pixel distance from LU5 to PC7 (the forearm length).
        
    \end{itemize}
    
    While RE shows error comparison across images of varying scales, AE provides a tangible metric for real-world application. We measured the actual physical length ${L}$ corresponding to the normalization constant ${D}$ for each subject. For instance, for the subject shown in \Cref{fig:acupoint sys and prof}, the physical distance from EX-UE10 to EX-UE11 is 4.9$\,\mathrm{cm}$, and from LU5 to PC7 is $25\,\mathrm{cm}$. Using the scale factor ${L/D}$, we can compute the actual error distance between all $\boldsymbol{P}_\text{pred}$ and $\boldsymbol{P}_\text{marked}$, i.e., AE.

    Determining the acceptable error range for an acupoint has been a longstanding challenge in TCM research due to the lack of definitive size standards~\cite{li2013big}. Previous research on automatic acupoint detection also lacks a commonly accepted evaluation method. To address this, we established a standardized criterion based on the distribution maps of acupoints marked by professional TCM practitioners.

    Drawing upon prior research in TCM that quantified the ``fuzziness" of acupoint localization, we observed that the distribution of labeled placements by experienced physicians forms an elliptical field rather than a single point~\cite{molsberger2012acupuncture}. To convert these clinical statistical observations into an engineering evaluation standard, we adopted a conservative approach. We selected the length of the short semi-axis—rather than the long semi-axis—as the radius to define a circular error tolerance range. This approach ensures that our evaluation criteria impose a stricter precision requirement that is valid in all directions, regardless of the ellipse's orientation.

    Utilizing the lengths statistical data of short semi-axis from these distribution maps, we defined two criteria based on confidence intervals:

    \begin{itemize}

        \item General Criterion (GC): Defined by the 95\% confidence interval, representing the effective therapeutic range of an acupoint.
    
        \item Precise Criterion (PC): Defined by the 63\% confidence interval, representing a narrower, highly accurate range.
        
    \end{itemize}

    Finally, we aggregated the data by body region to establish specific thresholds. By calculating the average short semi-axis lengths for representative acupoints in each region, we determined the final thresholds:

    \begin{itemize}

        \item PC: Hand ($0.56\,\mathrm{cm}$), Limbs ($0.94\,\mathrm{cm}$), Torso ($1.16\,\mathrm{cm}$).
    
        \item GC: Hand ($1.13\,\mathrm{cm}$), Limbs ($1.90\,\mathrm{cm}$), Torso ($2.14\,\mathrm{cm}$).
        
    \end{itemize}

The results are shown in \autoref{tab:ae average}. In a static state, the average AE for hand acupoint detection is $0.322\,\mathrm{cm}$, while for acupoints on the limbs and torso, the AE is $1.205\,\mathrm{cm}$ and $1.426\,\mathrm{cm}$, respectively.
When the body moves or changes angles, the accuracy decreases slightly due to tracking delays and joint recognition deviations. 
In acupoint therapy, different body parts require various levels of precision, with hand acupoints demanding higher accuracy than those on other body parts. 
Experimental results show that our system achieves higher accuracy than GC. Although its performance is slightly below PC, it remains adequate for educational purposes. 

The average AE of all detected acupoints under movement and rotation states is shown in \autoref{tab:ae acu}.

\vspace{-1em}
\begin{table}[h]
\caption{Average offset error and evaluation criteria.}
\label{tab:ae average}
\scriptsize
\centering
 \resizebox{\linewidth}{!}{
\begin{tabular}{cccccc}
\hline
\textbf{State}                         & \textbf{Regions} & \textbf{RE} & \textbf{AE} & \textbf{GC} & \textbf{PC} \\ \hline
\multirow{3}{*}{Movement and rotation} & Hands             & 0.104       & 0.519\,$\mathrm{cm}$     & -           & -          \\
                                   & Limbs             & 0.063       & 1.568\,$\mathrm{cm}$     & -           & -           \\
                                   & Torso            & 0.077       & 1.923\,$\mathrm{cm}$     & -           & -           \\ \hline
\multirow{3}{*}{Static}                & Hands             & 0.065       & 0.322\,$\mathrm{cm}$     & 1.13\,$\mathrm{cm}$      & 0.56\,$\mathrm{cm}$      \\
                                   & Limbs             & 0.048       & 1.205\,$\mathrm{cm}$     & 1.90\,$\mathrm{cm}$      & 0.94\,$\mathrm{cm}$      \\
                                   & Torso            & 0.057       & 1.426\,$\mathrm{cm}$     & 2.14\,$\mathrm{cm}$      & 1.16\,$\mathrm{cm}$      \\ \hline
\end{tabular}
}
\vspace{-1em}
\end{table}

\begin{table}[h]
  \caption{Average Absolute Error (AE) of all detected acupoints.}
  \label{tab:ae acu}
  \scriptsize
	\centering
    \resizebox{\linewidth}{!}{
\begin{tabular}{cccccccc}
\hline
\textbf{CV22} & \textbf{ST17} & \textbf{CV17} & \textbf{CV4} & \textbf{LU5} & \textbf{ST35} & \textbf{ST36} & \textbf{SI1} \\ \hline
$1.5\,\mathrm{cm}$ & $2.7\,\mathrm{cm}$ & $1.6\,\mathrm{cm}$ & $1.7\,\mathrm{cm}$ & $2.1\,\mathrm{cm}$ & $1.4\,\mathrm{cm}$ & $1.5\,\mathrm{cm}$ & $0.46\,\mathrm{cm}$ \\ \hline
\textbf{HT9} & \textbf{TE1} & \textbf{TE3} & \textbf{EX-UE8} & \textbf{EX-UE5} & \textbf{LI3} & \textbf{LI4} & \textbf{EX-UE9} \\ \hline
$0.58\,\mathrm{cm}$ & $0.43\,\mathrm{cm}$ & $0.46\,\mathrm{cm}$ & $0.47\,\mathrm{cm}$ & $0.95\,\mathrm{cm}$ & $0.46\,\mathrm{cm}$ & $0.61\,\mathrm{cm}$ & $0.54\,\mathrm{cm}$ \\ \hline
\textbf{EX-UE7} & \textbf{EX-UE4} & \textbf{EX-UE6} & \textbf{EX-UE11} & \textbf{EX-UE10} & \textbf{LU10} & \textbf{PC8} & \textbf{HT8} \\ \hline
$0.54\,\mathrm{cm}$ & $0.59\,\mathrm{cm}$ & $0.49\,\mathrm{cm}$ & $0.48\,\mathrm{cm}$ & $0.34\,\mathrm{cm}$ & $0.49\,\mathrm{cm}$ & $0.64\,\mathrm{cm}$ & $0.31\,\mathrm{cm}$ \\ \hline
\end{tabular}
}
\end{table}

For the test on different body types, two overweight individuals (B, C) and one tall individual (A) voluntarily participate, the average error results indicate that a taller stature has little impact on the acupoint detection error, but being overweight does increase the error, as illustrated in \autoref{tab:body type}. However, even for overweight individuals, the system's performance remains superior to the GC.

\begin{figure}[htb]
  \centering
  \includegraphics[width=1\linewidth]{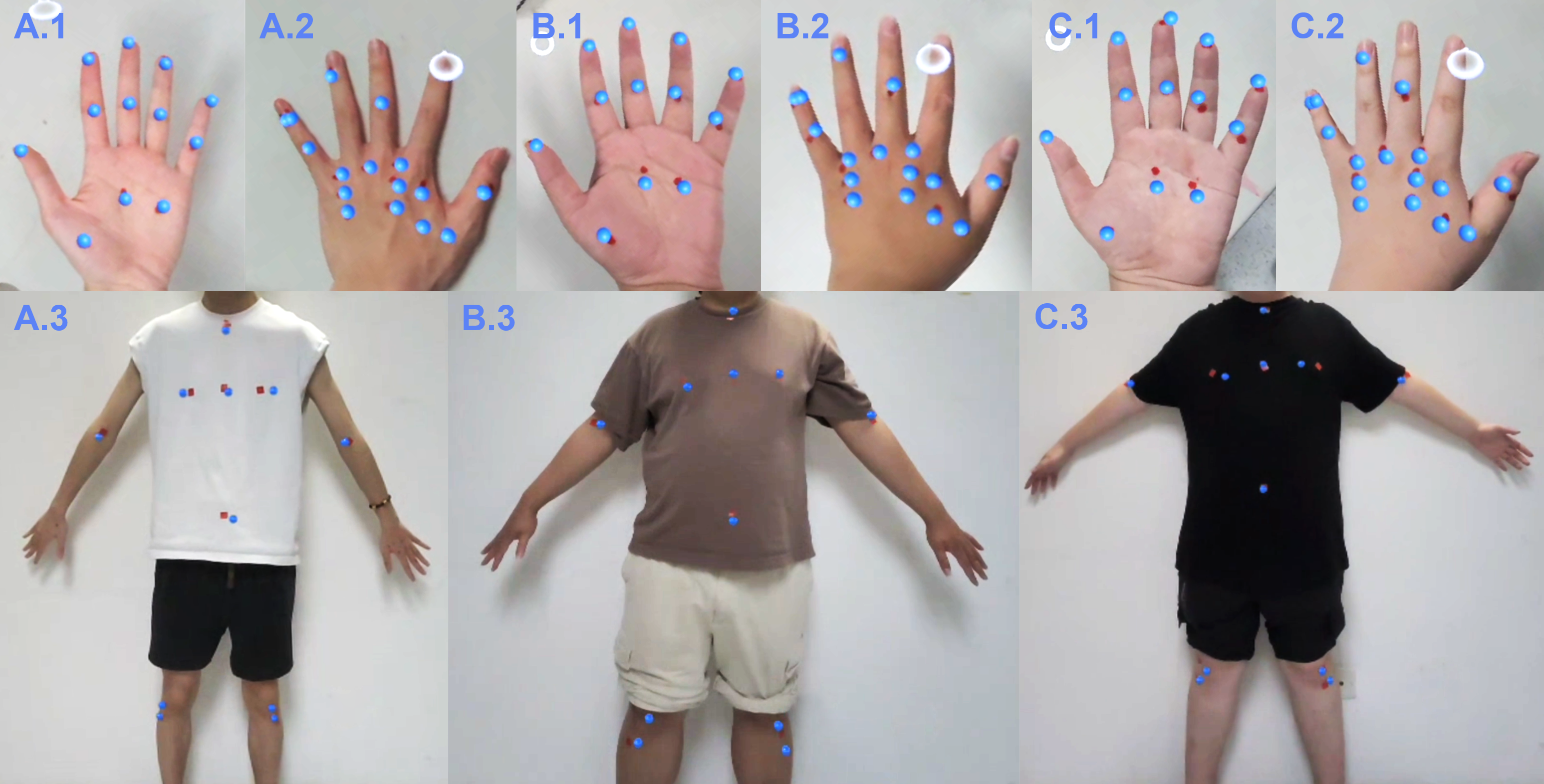}
  \caption{
Acupoints of individuals with different body types detected by our method (blue) and marked by the expert E1 (red).
  }
  \label{fig:body type}
\end{figure}

The method also works well on individuals with diverse body types, as shown in \Cref{fig:body type}. For the test on different body types, two overweight individuals (B, C) and one tall individual (A) voluntarily participate, the average error results indicate that a taller stature has little impact on the acupoint detection error, but being overweight does increase the error, as illustrated in \autoref{tab:body type}. However, even for overweight individuals, the system's performance remains superior to the GC. These results demonstrate the generality of the method across different individuals. 

\begin{table}[h]
\caption{Average offset error of different body types.}
\resizebox{\linewidth}{!}{
\renewcommand{\arraystretch}{1.2}
\begin{tabular}{cccccc}
\hline
\textbf{Body type} & \textbf{Average BMI} & \textbf{State} & \textbf{Regions} & \textbf{AE} & \textbf{GC} \\ \hline
\multirow{2}{*}{Overweight} & \multirow{2}{*}{30.34} & \multirow{2}{*}{Static} & Hands & $0.382\,\mathrm{cm}$ & $1.13\,\mathrm{cm}$ \\ 
 &  &  & Limbs and torso & $1.858\,\mathrm{cm}$ & $2.00\,\mathrm{cm}$ \\ \hline
\multirow{2}{*}{Tall} & \multirow{2}{*}{17.90} & \multirow{2}{*}{Static} & Hands & $0.323\,\mathrm{cm}$ & $1.13\,\mathrm{cm}$ \\ 
 &  &  & Limbs and torso & $1.395\,\mathrm{cm}$ & $2.00\,\mathrm{cm}$ \\ \hline
\end{tabular}
}
\label{tab:body type}
\end{table}

\subsection{Runtime Performance Analysis}
We evaluated MRATTS's runtime performance by analyzing its acupoint detection latency and the frame rate.
Concerning the latency of acupoint detection on the body (limbs and torso), we observed that the majority of the total latency (approximately 77.78\%) is attributed to video streaming, about $260\,\mathrm{ms}$. In contrast, the latency introduced by MRATTS accounts for only 11.63\%, with a total latency of $36.01\,\mathrm{ms}$.
Specifically, the MRATTS latency includes Mediapipe and SolvePnP ($19.15\,\mathrm{ms}$), WebSocket transmission ($5.44\,\mathrm{ms}$), B-cun calculation ($0.31\,\mathrm{ms}$), and rendering ($11.11\,\mathrm{ms}$). 
The visualization of acupoints achieves a stable frame rate of 90 FPS.
\newcolumntype{P}[1]{>{\RaggedRight\arraybackslash}p{#1}}

\section{User-Oriented Evaluation}
\label{sec:userStudies}
The effectiveness of MRATTS for acupoint therapy learning is evaluated through a controlled user experiment and expert feedback.

\subsection{Controlled User Study}
The controlled user study tested the learning effectiveness of 1. \emph{acupoint locations}, and 2. \emph{acupoint therapy knowledge}.
The study was conducted in our laboratory with the aforementioned setup (\Cref{sec:numericalStudy}).
\subsubsection{Study Design and Tasks}

A between-subjects design is used for the study. 
The \emph{acupoint localization} study tested the performance of three groups: the experimental MR Group (MG) using MRATTS on real human subjects; the active control VR Group (VG) using a prior method on the virtual standard acupoint model; and the baseline control Traditional Group (TG) using textbook-based methods. 
For the learning of \emph{acupoint therapy knowledge}, we compared two groups: the MG and the TG. 
The VG was excluded from this phase as its operational practice is similar to that of the MG.

We formulate our hypotheses for MG against other groups as follows. 
\begin{itemize}
    \item H1: MG participants are more proficient with acupoint locations.
    \item H2: MG participants have a better understanding of acupoint therapy knowledge.
    \item H3: Repeated simulated practice enables MG participants to progressively improve their operational skills.
    \item H4: MG participants report positive user satisfaction with MRATTS.
    \item H5: MG participants exhibit superior long-term knowledge retention.
\end{itemize}

To test these hypotheses, we formulated five tasks as summarized in \autoref{tab:tasks} and detailed in ~\Cref{sec:procedure}. 
The knowledge test questionnaire can be found in Appendix~\ref{supp.}.
\begin{table}[htb]
    \caption{Hypotheses and Corresponding Tasks of the User Study}
    \begin{tabularx}{\linewidth}{P{1cm}P{7cm}}
    \toprule
    Hypothesis &  Task              \\ \midrule
    H1                  & T1. Participants mark the locations of CV22, LU10, and LI4 on a real human subject after the learning session.                              \\
    H2                  & T2. After the learning session,   participants complete a knowledge test covering both theoretical and therapeutic concepts.                   \\ 
    H3                  & T3. Participants complete the simulated practice operations three consecutive times and review the user performance report after each trial.                 \\ 
    H4                  & T4. Participants complete a 7-point Likert scale questionnaire after the simulated practice~\cite{o2013examining}. \\ 
    H5                  & T5. Six months later, without prior review or additional learning, participants repeat the assessments defined in   T1 and T2. \\\bottomrule
    \end{tabularx}
    \label{tab:tasks}
\end{table}
    
\subsubsection{Participants}
Given the fundamental difference in information presentation between MR and traditional 2D media (textbooks), we anticipated a large effect size (Cohen's $d \geq 1.0$) for the primary measure of acupoint localization accuracy. 
The power analysis indicated that a sample size of 14 participants per group would be sufficient to achieve a statistical power of 0.80 ($\alpha = 0.05$) for detecting significant differences in pairwise comparisons. 
As participants are randomly assigned to three groups, the total number of participants is 42  ($N=14\times 3$).

Through invitations via social media, we successfully recruited 42 volunteers (21 male, 21 female) for the study. 
For all participants, we assessed their prior knowledge and experience before the experiment through a questionnaire. 
Only one participant had independently studied TCM. None of them had prior professional training in TCM. Regarding technology familiarity, only one participant was highly familiar with MR technology, while six had some familiarity with AR/VR.
Most participants had never or rarely used HMDs previously.

\subsubsection{Procedure}
\label{sec:procedure}
Each participant was randomly assigned to the MG, VG, or TG. 
Then, we introduced the objectives of the study to all participants and provided them with group-specific task sheets. 
After obtaining informed consent, a 30-minute pre-experiment preparation session was conducted. 

Participants in MG and VG were trained to use PICO 4 Pro, including basic gesture interactions (e.g., clicking, dragging) and the pinch interaction of virtual needles and moxa sticks. 
We also demonstrated the user interface and operational procedures of MRATTS to the participants.

Next, the MG participants utilized the acupoint detection module in MRATTS to learn acupoint locations on a real person in MR, while the VG participants used a virtual standard acupoint model, and the TG participants relied on textbooks. 
All groups had 5 minutes to learn about 8 acupoints (three of which were selected for testing). 
Upon completion of the learning sessions, participants from all three groups were required to mark the acupoints on the same individual within 5 minutes (\textbf{T1}). The offset distance between the marked locations and those indicated by the professional was calculated and used to analyze the differences among the groups.

Subsequently, we informed the participants of the specific ranges of acupoint therapy knowledge involved in the experiment. 
The MG utilized the simulated acupoint therapy practice module in MRATTS to perform each involved operation three consecutive times (\textbf{T3}). 
The TG relied on textbooks (with the aforementioned learning scope highlighted). Both groups were given 10 minutes for practice or study. Following their respective sessions, both the MG and TG completed a knowledge test questionnaire (\textbf{T2}) within 5 minutes.

After the learning phase, we used a 7-point Likert scale questionnaire with 20 questions to investigate the experience of participants within 10 minutes (\textbf{T4}). Additionally, we conducted post-experiment interviews with the participants to gather qualitative feedback regarding MRATTS.

Six months later, we re-invited the original participants ($N=40$, two were unable to participate due to personal reasons) to repeat the acupoint marking task and the knowledge test questionnaire (\textbf{T5)} to assess long-term memory retention.

 \begin{figure*}[tbh]
      \centering
      \includegraphics[width=1\linewidth]{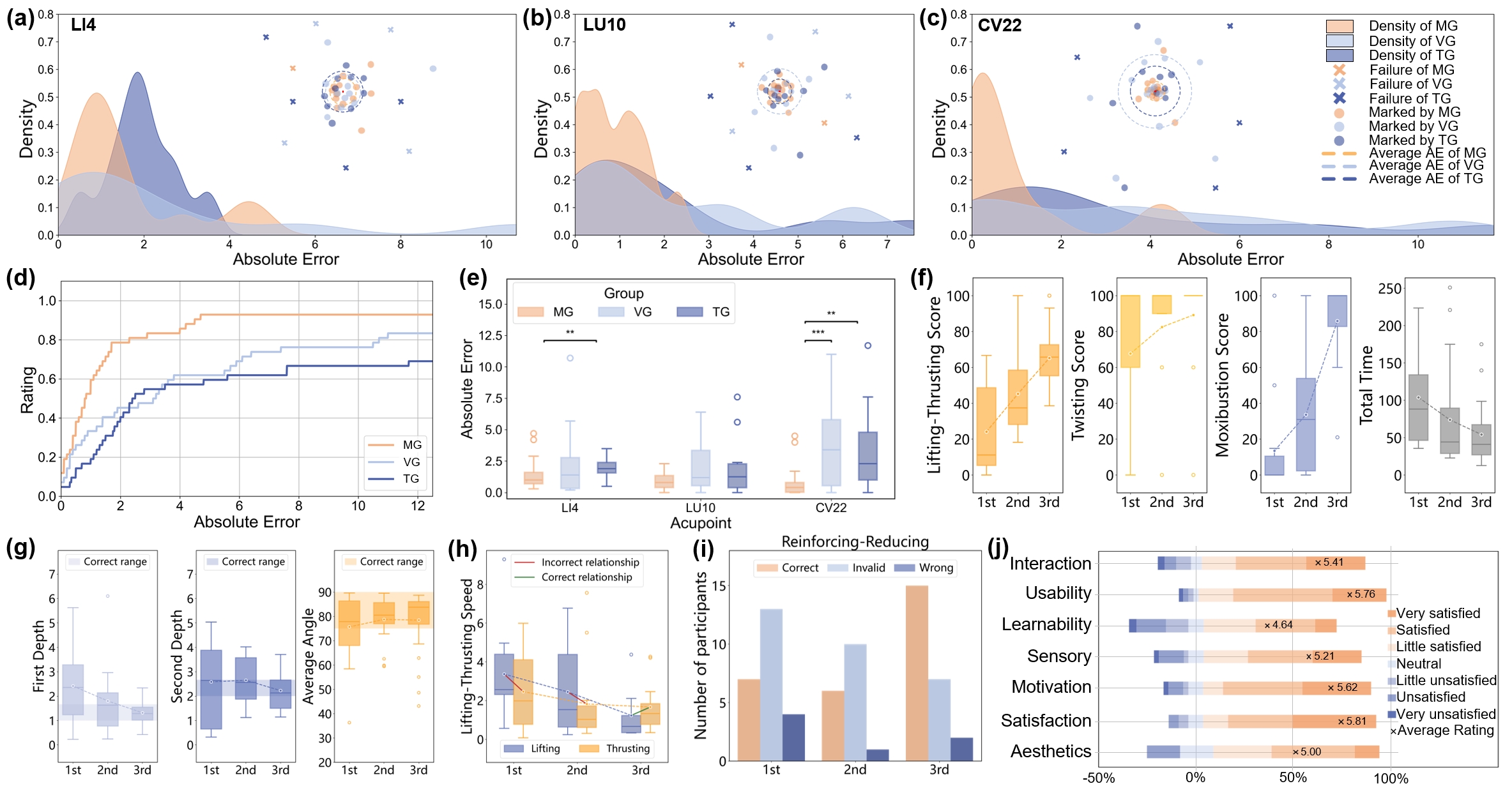}
      \caption{Results of the user study.  
      (a--c) AE of the acupoint marked by MG, VG, and TG on a real person. The density plots show differences in the distribution and central tendency of AE among various groups. The scatter plots show individual AE values across groups and differences in their average AE. 
      (d) Accuracy curves with different AE thresholds. (e) Boxplots of AEs at acupoints LI4, LU10, and CV22 after learning. (f) Boxplots of scores and response time with the number of trials, (g) boxplots of depth and angle deviated from the correct ranges, (h) changes in lifting and thrusting speeds, and (i) the correct proportions of reinforcing and reducing states in the initial and final operations. (j) The questionnaire results of all MG members in Likert scales. 
    }
      \label{fig:T3}
\end{figure*}

\subsubsection{Study Results} 
The results of our user study are presented according to the five experimental tasks T1---T5.

\noindent\textbf{T1: Acupoint localization learning.}

We evaluated the acupoint localization ability of participants with different learning methods by comparing the markers of participants with those of E1 and E2.
Markers far from the correct anatomical region were considered failures. 
The MG achieved the highest success rate ($92.86\%$), followed by the VG ($83.33\%$) and TG ($69.05\%$).  Despite these markings being successful, differences in accuracy of location remained evident among the groups.

Scatter plot analysis indicated that the MG outperformed both VG and TG in terms of Average AE, whereas the VG and TG showed varying strengths and weaknesses across different acupoints. Analysis of data distribution characteristics using a density plot revealed that the MG exhibited the most concentrated data distribution, with the smallest error distance at the peak, showing notably superior overall performance compared to the VG and TG. Although the VG demonstrated a slightly smaller peak error distance than the TG, its overall data distribution was more dispersed, indicating a less concentrated distribution, as shown in main paper \Cref{fig:T3}(a--c).

From the box plot in \Cref{fig:T3}(e) of the main paper, we observed a similar trend: the MG exhibited the smallest error range and the highest stability. Although the VG and TG each showed strengths in terms of median and upper quartile values, the overall stability of the VG was significantly lower. After combining statistical outliers with marking failures, the success rates of each group were: 83.33\% for the MG, 80.95\% for the VG, and 66.67\% for the TG.

We further plotted accuracy curves for each group by setting different AE values as thresholds, as depicted in \Cref{fig:T3}(d) of the main paper. When the accuracy of the MG stabilized around 90\% (at a threshold of $4.7\,\mathrm{cm}$), the accuracies for the VG and TG reached only approximately 60\%. Using 80\% accuracy as the teaching standard, the MG required an error threshold of only $2.3\,\mathrm{cm}$, whereas the VG had to extend this threshold to $10.7\,\mathrm{cm}$, and the TG failed to achieve this standard entirely. Although the overall accuracy of the VG was higher than that of the TG, its performance was slightly inferior within the $2 \text{--} 4\,\mathrm{cm}$ error range.

These results demonstrate that MRATTS is superior to 3D models or traditional textbooks for learning both the anatomical regions and the accurate locations of acupoints. While learning with 3D models showed slight advantages over textbooks in location accuracy, it exhibited notably poorer stability and general applicability. A detailed analysis of these findings is as follows: 

\begin{itemize}
    \item (1) MG: Participants used MRATTS  to dynamically observe acupoints directly on a real person, allowing them to combine real-time observation with authentic skin textures as a reference, thus making precise localization easier and enhancing memory retention. 
    \item (2) VG: Although the 3D model offered some advantages in spatial perception and memorization, the absence of realistic skin texture details and variations in participants' acceptance of the model resulted in unstable learning outcomes. 
    \item (3) TG: Although participants using traditional textbooks tended to forget acupoint locations more easily, this method had higher general applicability. 
Textbook illustrations highlight skin surface landmarks and texture lines. 
Despite inherent ambiguities in 2D images, they provide greater stability for learning acupoint locations.
\end{itemize}

\noindent\textbf{T2: Acupoint therapy knowledge learning.}

The questionnaire result reveals a significant improvement in acupoint therapeutic knowledge for the MR group ($p = 0.046 < 0.05$), Cohen’s $d = 0.793$, suggesting that simulated practice in MR enhances learning compared to traditional methods. 
However, no significant difference was found for theoretical knowledge learning ($p = 0.868$), as shown in \autoref{tab:T} and \autoref{tab:U}. 

\begin{table}[h]
\vspace{-1em}
\caption{The result of Independent Samples T-Test on acupoint therapy knowledge}
\renewcommand\arraystretch{1}
\resizebox{\linewidth}{!}{
\begin{tabular}{ccccc}
\hline
\textbf{Group} & \textbf{Mean} & \textbf{Standard deviation} & \textbf{\textit{p}}               & \textbf{Cohen's \textit{d}}     \\ \hline
MG             & 6.500         & 2.103       & \multirow{2}{*}{0.046**} & \multirow{2}{*}{0.793} \\
TG             & 5.071         & 1.439       &                          &                        \\ \hline
\end{tabular}   
\label{tab:T}
}
\vspace{-1em}
\end{table}

\begin{table}[h]
\vspace{-1em}
\caption{The result of Mann-Whitney U test on acupoint theoretical knowledge}
\renewcommand\arraystretch{1}
\resizebox{\linewidth}{!}
{
\begin{tabular}{ccccc}
\hline
\textbf{Group} & \textbf{Mean} & \textbf{Standard deviation} & \textbf{\textit{p}}               & \textbf{Cohen's \textit{d}}     \\ \hline
MG             & 2.071         & 1.207         & \multirow{2}{*}{0.868} & \multirow{2}{*}{0.059} \\
TG             & 2.143         & 1.231        &                          &                         \\ \hline
\end{tabular}
}
\vspace{-1em}
\label{tab:U}
\end{table}

We attribute this to the fact that learning and memorizing foundational knowledge primarily relies on textual information, and there is no substantial difference between the textual content presented in MR and in textbooks.

\noindent\textbf{T3: Improvement in simulation practice.}

We analyze the scores of simulated practices and the detailed data of each process. 
We find that as the number of experiments increases, the operation time of the participants becomes shorter---participants became more proficient in the operation of MR, and they could complete the corresponding task more quickly. 
Meanwhile, there is an increasing trend in the scores of acupuncture lifting-thrusting, twisting, and moxibustion, as shown in~\Cref{fig:T3}(f). 
This suggests that the evaluation and visual data feedback during the practice may be helpful for learning.

As in~\Cref{fig:T3}(g), it is evident that during the initial insertion operation, users often penetrate too deeply, but by the third trial, they can correct this error and insert within the correct range. 
Similarly, the speed of lifting-thrusting differs between the reinforcing and reducing techniques. 
In the first attempts, the average lifting-thrusting speed is incorrect. 
A proper reinforcing technique requires fast thrusting and slow lifting. 
We observed that with more training, users gradually learned the correct lifting-thrusting speed: by the third attempt, their performances improved noticeably, as shown in~\Cref{fig:T3}(h). 
Regarding the completion of reinforcing-reducing operations (\Cref{fig:T3}(i)), initial attempts of participants were ineffective because both reinforcing and reducing should satisfy conditions of speed and depth. 
However, in the final trial, $62.5\%$ of the reinforcing-reducing types were performed correctly---an increase of $33\%$. 
The proportion of incorrect types dropped to $8.3\%$, and the percentage of ineffective types decreased by $25\%$.

\noindent\textbf{T4: User experience with MRATTS.}

Most responses fall on the positive side of the Likert scale, indicating overall user satisfaction with the system, as shown in \Cref{fig:T3}(j). 
Among the seven evaluated aspects, usability scored the highest, while ease of learning scored the lowest, consistent with the longer completion time observed in T3, suggesting a learning curve for MRATTS. 
A few participants gave interaction, usability, and sensory experience relatively low ratings, possibly due to limitations in MR performance. 
The lack of a depth sensor in PICO 4 Pro affects occlusion perception, reducing immersion. 
Hand gesture recognition can be compromised by occlusions, and the deep integration of TCM theory may create a sense of unfamiliarity for some users.

\noindent\textbf{T5: Knowledge retention test.}

In terms of learning acupoint locations, the performance of all learners declined after a six-month memory retention period.  The success rate for all groups dropped significantly, and the average error distance (calculated only from correctly recalled data) for MG and VG increased, as illustrated in \autoref{tab:location remember}.

However, it is noteworthy that the average error distance for TG actually decreased after six months. An in-depth analysis of individual data suggests this is a statistical artifact attributable to ``survivorship bias". The textbook-based learning method led to polarized outcomes: learners who were initially less certain (high-error) tended to forget the locations completely after six months, while those who were initially precise maintained their accuracy.  As a result, the high-error participants were naturally filtered out of the error calculation, which in turn lowered the group's overall average error distance.

This highlights the need to analyze both accuracy and error rates in conjunction.  In the MRATTS-based approach, learners initially developed a more uniform level of precision.  Six months later, despite a natural decline in memory and precision, our method remained superior to the traditional approaches on both metrics.

\begin{table}[h]
\caption{Knowledge retention test result of acupoint location learning.}
\resizebox{\linewidth}{!}{
\begin{tabular}{ccccc}
\hline
\multirow{2}{*}{\textbf{Group}} & \multicolumn{2}{c}{\textbf{Post-learning}}      & \multicolumn{2}{c}{\textbf{6-Month Follow-up}}  \\
                                & \textbf{Success rate} & \textbf{Error distance} & \textbf{Success rate} & \textbf{Error distance} \\ \hline
MG                              & 92.86\%               & $1.16\,\mathrm{cm}$                & 72.20\%               & $1.38\,\mathrm{cm}$                \\
VG                              & 83.33\%               & $2.99\,\mathrm{cm}$                & 45.24\%               & $3.01\,\mathrm{cm}$                \\
TG                              & 69.05\%               & $2.50\,\mathrm{cm}$                     & 38.89\%               & $1.57\,\mathrm{cm}$           \\ \hline  
\end{tabular}
}
\label{tab:location remember}
\end{table}

After six months, there were no significant differences between the two groups in their Theoretical scores, Therapeutic scores, or total scores, as shown in \autoref{tab:knowledge remember}. It should be noted that, in terms of acupoint theoretical knowledge, the average score for TG decreased by 11.24\% over the six-month period, while the MG's score decreased by 20.51\%. Although the MRATTS-based method can improve learning efficiency and result in higher scores immediately after learning, it exhibits some shortcomings in long-term memory retention compared to traditional methods.

\begin{table}[h]
\caption{Knowledge retention test result of acupoint theoretical and therapeutic knowledge learning.}
\resizebox{\linewidth}{!}{
\begin{tabular}{lcccccc}
\hline
\multicolumn{1}{c}{\multirow{2}{*}{\textbf{Group}}} & \multicolumn{3}{c}{\textbf{Post-learning}}                                                                                                 & \multicolumn{3}{c}{\textbf{6-Month Follow-up}}                                                                                             \\
\multicolumn{1}{c}{}                                & \multicolumn{1}{l}{\textbf{Theoretical}} & \multicolumn{1}{l}{\textbf{Therapeutic}} & \multicolumn{1}{l}{\textbf{Total}} & \multicolumn{1}{l}{\textbf{Theoretical}} & \multicolumn{1}{l}{\textbf{Therapeutic}} & \multicolumn{1}{l}{\textbf{Total}} \\ \hline
MG                                                  & 2.07                                           & 6.50                                           & 8.57                                     & 1.50                                           & 5.17                                           & 6.67                                     \\
TG                                                  & 2.14                                           & 5.07                                           & 7.21                                     & 1.67                                           & 4.50                                           & 6.17                                     \\ \hline
\end{tabular}
}
\label{tab:knowledge remember}
\end{table}

\vspace{-1em}
\subsection{Expert Feedback}
The expert E1 evaluated both MRATTS and the experimental procedure. The entire expert review session lasted approximately 20 minutes. 
After informed consent, we provided an overview of the functions of MRATTS. Subsequently, the expert assessed the pedagogical feasibility of the acupoint detection and simulated therapy practice modules. 
Then, E1 reviewed our experimental workflow. The feedback provided valuable insights into the educational utility and clinical potential of MRATTS.

The expert review offered a professional assessment of the strengths and limitations of MRATTS from a clinical perspective. 
First, the expert acknowledged that the experimental design of the study is comprehensive. 
Regarding the performance of MRATTS, E1 noted that the acupoint detection accuracy satisfies teaching requirements. Furthermore, the expert stated, ``\emph{MRATTS helps learners understand anatomical terms and facilitates memory retention.}"
However, E1 pointed out that for certain acupoints relying more on touching surface landmarks for determination (such as ST17), the performance is slightly weaker, emphasizing that ``\emph{palpation is still essential for precise localization.}" 
Despite this limitation, E1 emphasized that the acupoint therapy operations provide an intuitive teaching method for entry-level learners with limited prior knowledge, thereby improving learning efficiency. 
Overall, the expert feedback validates MRATTS as a promising supplementary tool for modern TCM education.

\section{Discussion}
\label{sec:discussion}

\subsection{Hypotheses Validation}
Results of the localization task (T1) confirm that the MG significantly outperformed both control groups in marking success rate. Furthermore, density and scatter analyses reveal that MG participants achieved not only the lowest average Absolute Error but also the highest stability in performance compared to the VG and TG (H1---supported). 

The MG showed significant improvement in therapeutic knowledge compared to the TG. However, no significant difference was found in theoretical scores. The approach of separating the knowledge test into theoretical and therapeutic dimensions yielded results that differ from prior findings~\cite{sun2023design}. This suggests that MR visualization specifically benefits the comprehension of complex operational techniques rather than the memorization of text-based theory (H2---partially supported).

Operational data from T3 demonstrates a consistent trend of decreased operation time and increased proficiency scores across the three trials. The progressive correction of lifting-thrusting speeds and reinforcing-reducing sequences confirms the efficacy of the system’s real-time visual guidance (H3---supported).

Positive responses on the Likert scale (T4), particularly regarding usability, indicate high user satisfaction. This suggests that MRATTS effectively makes complex TCM techniques accessible and engaging for learners (H4---supported).

The six-month follow-up data (T5) indicate that while MR-based learning led to significantly better long-term retention of spatial acupoint locations, the retention of general theoretical and therapeutic knowledge did not differ significantly from traditional methods (H5---partially supported).

\subsection{Limitations} 
Our sample size ($N=42$) was determined based on an anticipated very large effect size ($d \ge 1.0$).
While T1 results support this assumption ($d$ ranging from 0.966 to 1.118), the simulated practice (T3) only yields a large effect size ($d=0.793$). 
This indicates that while the study was sufficiently powered for localization tasks, it may be statistically underpowered for detecting subtler variations in complex therapeutic skills.
The knowledge retention analysis is limited to a six-month follow-up point. A more robust evaluation would incorporate more frequent time points (e.g., at three months) to better map the long-term memory retention curves for each learning method.

Another limitation is the smoothness and accuracy of acupoint tracking. 
Acupoint tracking during the movement of a subject is not as smooth as for still poses. It is because the tracking relies on video streaming to a server, which introduces stream latency. 
Future work should explore on-device processing to reduce this latency.
Tracking accuracy is not uniform across the body, with lower precision on the torso and limbs than hands; for hands, the accuracy depends on a specific position relative to the cameras of the HMD. 
In terms of detection accuracy evaluation, our ground truth was established by an experienced professional without an inter-rater reliability analysis. 
Furthermore, hand-interaction occlusions can cause detection jitter.

\section{Conclusion and Future Work}
This paper introduces MRATTS, an MR-based TCM acupoint therapy teaching system with three key features: real-time acupoint detection and visualization in MR, simulated practice of advanced acupoint therapy with real-time guidance, and a set of evaluation standards with a user performance report. 
With MRATTS, users can practice acupressure, acupuncture, and moxibustion based on real-time acupoint detection, and subsequently improve their learning accuracy and efficiency using the scores and data charts provided in the performance report. 
Through numerical and user-oriented evaluations, the effectiveness and usefulness of MRATTS are demonstrated.

For future work, we would like to provide a more realistic user experience by improving hand interactions and integrating tactile feedback~\cite{yu2020fully,Moon2023,Shi2024}. 
We would also like to leverage advanced information visualization and visual guidance~\cite{pietschmann2023quantifying}.
Finally, it is also in our interest to create a standardized patient to simulate symptoms for more advanced and realistic training.

\section*{Acknowledgments}

We wish to thank our collaborators Chengkai Zou for his assistance with the illustrations in this paper and Junhao Xie for his help in conducting the user study.

 \bibliography{ref}
 \bibliographystyle{IEEEtran}

\appendices
\twocolumn[\section{Knowledge Test Questionnaire}\label{supp.}]

\subsection*{I. Acupoint Location:}
\begin{enumerate}[label=\arabic*.]
    \item Hegu (LI4)
    \vspace{-0.1em}
    \item Yuji (LU10)
    \vspace{-0.1em}
    \item Tiantu (CV22)
\end{enumerate}

\subsection*{II. Acupoint Theoretical Knowledge:}
\begin{enumerate}[label=\arabic*.]
    \item Which hand acupoint treats cough?
    
    \begin{itemize}
        \item[ ] A. Laogong (PC8) 
        \item[ ] B. Sifeng (EX-UE10) 
        \item[ ] C. Yuji (LU10) 
        \item[ ] D. Dubi (ST35)
    \end{itemize}
    
    \item Which is a Pericardium Meridian acupoint?
    
    \begin{itemize}
        \item[ ] A. Jiexi (ST41) 
        \item[ ] B. Hegu (LI4) 
        \item[ ] C. Tiantu (CV22) 
        \item[ ] D. Laogong (PC8)
    \end{itemize}
    
    \item What is ST35?
    
    \begin{itemize}
        \item[ ] A. Tiantu (CV22) 
        \item[ ] B. Dubi (ST35) 
        \item[ ] C. Zusanli (ST36) 
        \item[ ] D. Jiexi (ST41)
    \end{itemize}
    
    \item What does ``CV" stand for in meridian names?
    
    \begin{itemize}
        \item[ ] A. Conception Vessel (Ren Mai) 
        \item[ ] B. Lung Meridian of Hand-Taiyin
        \item[ ] C. Governor Vessel (Du Mai) 
        \item[ ] D. Large Intestine Meridian of Hand-Yangming
    \end{itemize}
    
    \item Which stomach meridian acupoint treats headaches?
    
    \begin{itemize}
        \item[ ] A. Jiexi (ST41) 
        \item[ ] B. Tiantu (CV22) 
        \item[ ] C. Sifeng (EX-UE10) 
        \item[ ] D. Hegu (LI4)
    \end{itemize}
\end{enumerate}

\subsection*{III. Acupoint Therapy Knowledge:}
\begin{enumerate}[label=\arabic*.]
    \item Which angle is considered a ``subcutaneous insertion" in acupuncture?
    
    \begin{itemize}
        \item[ ] A. 0° 
        \item[ ] B. 15° 
        \item[ ] C. 30° 
        \item[ ] D. 45° 
        \item[ ] E. 90° 
    \end{itemize}
      
    \item For a patient with facial redness and restlessness, which technique should be used? 
    \begin{itemize}
        \item[ ] A. Reinforcing 
        \item[ ] B. Reducing
    \end{itemize}
    
    \item In lifting-thrusting reinforcing/reducing, which actions are for reinforcing? (Multiple choice)
    
    \begin{itemize}
        \item[ ] A. Shallow first, then deep
        \item[ ] B. Deep first, then shallow
        \item[ ] C. Rapid insertion, slow withdrawal (heavy insertion, light withdrawal)
        \item[ ] D. Slow insertion, rapid withdrawal (light insertion, heavy withdrawal)
    \end{itemize}
    
    \item In twisting needle techniques, which actions are for sedation? (Multiple choice)
    
    \begin{itemize}
        \item[ ] A. Clockwise first, then counterclockwise 
        \item[ ] B. Counterclockwise first, then clockwise
        \item[ ] C. Fast clockwise, slow counterclockwise 
        \item[ ] D. Fast counterclockwise, slow clockwise
    \end{itemize}
    
    \item Standard moxibustion distance from the skin:
    
    \begin{itemize}
        \item[ ] A. $1\,\mathrm{cm}$ 
        \item[ ] B. $3\,\mathrm{cm}$ 
        \item[ ] C. $5\,\mathrm{cm}$ 
        \item[ ] D. $7\,\mathrm{cm}$
    \end{itemize}
    
    \item Correct needle depth for Yuji (LU10) ($1\,\mathrm{cun}$ $\approx$ $3.33\,\mathrm{cm}$):
    
    \begin{itemize}
        \item[ ] A. $0.2\,\mathrm{cun}$ 
        \item[ ] B. $0.7\,\mathrm{cun}$ 
        \item[ ] C. $1.2\,\mathrm{cun}$ 
        \item[ ] D. $1.5\,\mathrm{cun}$
    \end{itemize}
    
    \item Briefly describe the ``Sparrow-pecking" moxibustion procedure:
    
    \underline{\hspace{8cm}}
\end{enumerate}

\subsection*{IV. Open-Ended Questions:}
\begin{enumerate}[label=\arabic*.]
    \item What improvements would you suggest for our system?
    
    \vspace{2cm}
    
    \item What aspects of our system do you think deserve recognition?
    
    \vspace{2cm}
\end{enumerate}

\vfill

\end{document}